\documentclass{ws-rv9x6}
\usepackage{subfigure}   
\usepackage{ws-rv-thm}   
\usepackage[square]{ws-rv-van}   
\usepackage{bm}
\allowdisplaybreaks
\makeindex

\begin{document}

\chapter[Using World Scientific's Review Volume Document Style]{Chiral Electroweak Currents in Nuclei}\label{ra_ch1}

\author{D.O.\ Riska}

\address{Finnish Society of Science and Letters, Helsinki, Finland, and\\
The Cyprus Institute, Nicosia, Cyprus}

\author[D.O.\ Riska and R.\ Schiavilla]{R.\ Schiavilla}

\address{Theory Center, Jefferson Lab, Newport News, Virginia, and\\
Physics Department, Old Dominion University, Norfolk, Virginia}
\begin{abstract}
The development of the chiral dynamics based description of nuclear electroweak currents is reviewed. Gerald E. (Gerry) Brown's role in basing theoretical nuclear physics on chiral Lagrangians is emphasized.  Illustrative examples of the successful description of electroweak observables of light nuclei obtained from chiral
effective field theory are presented. 
\end{abstract}
\body


\section{Introduction}
The phenomenological success of the systematic application
of chiral effective field theory ($\chi$EFT) to the electromagnetic
and weak observables of light nuclei in the mass range $A=2$--10
has been remarkable. Here the background and early application
of chiral Lagrangians to nuclear current operators, and Gerry
Brown's role as an initiator of this approach are reviewed in the
next section. The third and fourth sections contain a summary of the
present stage of the $\chi$EFT approach and a set of illustrative
examples of its application to nuclear electroweak observables.

\section{Historical Perspective}
\label{history}

\subsection{Gerry Brown's early work with chiral Lagrangians}

Gerry Brown was one of the first physicists to appreciate the utility of effective chiral
Lagrangians in theoretical nuclear physics. In the late 1960s he planned
a major effort to derive a realistic nucleon-nucleon ($NN$) interaction model from
meson exchange, as he saw that the pair suppression built into Weinberg's non-linear
chiral Lagrangian for the pion-nucleon interaction~\cite{Wei} might solve the
over-binding problem, which the two-pion exchange interaction described with
the conventional pseudoscalar pion-nucleon coupling model inevitably led to.
That a realistic description of the nucleon-nucleon amplitude could be constructed
in this way was then demonstrated by Brown and Durso~\cite{BDur} and Chemtob,
Durso, and Riska~\cite{CDR}.

In parallel with this development Chemtob and Rho derived expressions for the
exchange current contributions to the electromagnetic and axial two-nucleon current
operators that arise from the effective chiral Lagrangians for pion and vector meson
exchange~\cite{CRho}. The $\rho$-meson exchange interaction complements the pion
exchange one by counteracting the strong tensor component of the latter and
improving the interaction models of Refs.~\cite{BDur,CDR}.  Concurrently, it enhances
the effects of the long range electromagnetic pion exchange current.

Gerry Brown's interest in the role of exchange currents in nuclei was stimulated by
Chemtob and Rho's estimate of the axial exchange current enhancement of the
Gamow-Teller matrix element in Tritium $\beta$-decay~\cite{CRho69}.  He suggested
that the D-state components in the trinucleon wave functions, even if small, could
enhance the calculated value. This was illustrated with a schematic model for the those
components~\cite{RisBro}. 

The axial exchange current is related to the $NN$ interaction indirectly
through the partially-conserved axial current (PCAC) relation to the pion-production
operator. The form of the electromagnetic exchange current is in contrast directly
constrained by the $NN$ interaction through the continuity equation~\cite{Ris85b,GRis}.
The first demonstrations that electromagnetic exchange currents could play a
significant role in nuclear observables were in fact related to those.  In 1971 Gerry
Brown's attention was drawn to the fact the well measured total cross section for
capture of thermal neutrons on protons---the process $^1{\rm H}(n,\gamma)^2{\rm H}$---could
not be fully explained by  the sum of the neutron and proton magnetic moments.
He asked one of us (DOR) to take into account the pion exchange current operators
derived by Chemtob and Rho~\cite{CRho}.  The result was that the exchange current
contribution can account for the $\sim10$\% difference between the calculated and the
experimental value for the cross section~\cite{RisBro1}.  The key part of the pion
exchange current operator was related to the chiral Lagrangian for the
pion-nucleon interaction.  A smaller, nevertheless significant, effect was due
to the pion exchange operator, which involved intermediate $\Delta$
resonances~\cite{Stran}.

\subsection{The sequel}

It was soon afterwards shown that the experimental values of the
$^3$H and $^3$He magnetic moments could also be almost fully
accounted for in the same way with realistic wave functions~\cite{Harp}.
Moreover it was found that the pion exchange current
contribution could provide about one half of the cross section for capture
of thermal neutrons on $^2$H~\cite{Hadj}.  Later it was shown that the exchange
current contribution to the calculated cross section for thermal neutron capture
on $^3$He is about 5 times larger than that from the sum of the three nucleon
magnetic moments~\cite{TowKha}.  Finally it was shown that the cross section
for radiative neutron capture on $^3$He is almost totally due to the exchange
current contribution~\cite{CarlsRSW}.

While the strength of the $NN$ interaction scales with the mass of
the exchanged system, the meson exchange magnetization operator scales
with the inverse mass of the exchanged system~\cite{DOR96}.  This makes
the relative contribution of short-range mechanisms to the matrix elements
of the exchange current operators much weaker than to the matrix elements
of the interaction.  The calculated cross section for backward electro-disintegration
of the deuteron, which is very strongly dependent on the electromagnetic
exchange currents, illustrates this well~\cite{Hockert}.  In this reaction the
matrix element of the single-nucleon current operators changes sign at fairly
low momentum transfer and therefore the cross section near that zero is built
up entirely from the exchange current contribution.  In this case the cross section
obtained with the pion-exchange current alone is quite similar to that calculated
later with wave functions that are consistent with a realistic phenomenological
interaction model~\cite{SchiaRis}.  In the case of the magnetic form factors of the
trinucleons there is a similar destructive interference between the matrix elements
of the single-nucleon current operators for the S- and D-state components of the wave
functions~\cite{BranKim}, with the consequence that the exchange current
contribution is very large~\cite{BarrHadj}.  With only the single-nucleon current
operator, the calculated magnetic form factors of the trinucleons would have zeroes
at fairly low values of momentum transfer, in disagreement with experiment.  This
has later been demonstrated with improved wave functions and exchange current
operators that satisfy the continuity equation with realistic interactions~\cite{ScPanda}.
In larger nuclei the main features in elastic and transition electromagnetic
form factors are due to the shell structure.  Even so, it has been shown that in
the case of Li, the inclusion of the exchange current contribution does markedly
improve the agreement with experiment for these form factors, once the interaction
currents are consistent with wave functions corresponding to realistic interaction
models~\cite{WirSch}.  The effect is smaller in $^{12}$C$\,$~\cite{LumbChe}
and in $^7$Li, $^9$Be and $^{10}$B~\cite{Duba}.

In the examples above, it is the isovector part of the pion exchange current
operator, which is numerically most important.  In the case of the magnetic
form factor of the deuteron, only the isoscalar part of the pion exchange
current contributes, and the most important term in that operator
involves a $\rho\pi\gamma$ transition. Without this exchange current operator
 the calculated magnetic moment of the deuteron would have a node close to
 momentum transfer values $\sim 6$ GeV/c, in disagreement with
 experiment~\cite{GariHyu,SchiaRis}.  Since the $\rho\pi\gamma$ exchange
 current is transverse, its form is not constrained by the $NN$ interaction
 by the continuity equation.  Its longer range form can however be determined
 in the Skyrme model approach by the chiral anomaly~\cite{NymRis,WakWei}.

While the two-nucleon exchange current operators give large contributions
to nuclear electromagnetic observables, it has been demonstrated that the
three-nucleon exchange current operators that are associated with two-pion
exchange with pion scattering off an intermediate nucleon, give but very small
contributions to the magnetic form factors of the trinucleons~\cite{SchiMarR}.
The remarkably successful nuclear phenomenology based on the simple
pion exchange operators that are related to the lowest order chiral Lagrangian for
the pion-nucleon interaction has later been explained within the context of
$\chi$EFT~\cite{ParkMinRho}.

\subsection{The axial exchange current}

The role of the pion exchange axial exchange current was first considered
for the case of the Gamow-Teller transition in the $\beta$-decay of Tritium.
Those results were reaffirmed with more accurate wave functions~\cite{FishKim}.
Gari and Huffman noted that this axial exchange current also contributes
a small enhancement of the cross section for the basic solar burning
reaction $^1{\rm H}(p,e^+\, \nu_e)^2{\rm H}$~\cite{GariHuff}.   This was
confirmed by Dautry, Rho, and Riska in a study of muon absorption in
the deuteron $^2{\rm H}(\mu^-,\nu_\mu)\,nn$ with an improved version
of the axial exchange current operator, which was checked for consistency
against the P-wave piece of the cross section for the reaction
$^1{\rm H}(p, \pi^+)^2{\rm H}$ near threshold~\cite{DautryRR}.
The same Hamiltonian was then used to calculate the matrix
elements for the solar neutrino processes $^1{\rm H}(p,e^+\, \nu_e)^2{\rm H}$
and $^1{\rm H}(p\, e^-, \nu_e)^2{\rm H}$ as well. 
A later calculation, with wave functions obtained from the realistic Argonne
$v_{14}$ $NN$ interaction~\cite{AV14} of the weak proton capture reactions
$^1{\rm H}(p,e^+\, \nu_e)^2{\rm H}$ and $^3{\rm He}(p,e^+\, \nu_e)^4{\rm He}$
normalized axial exchange current operator against the known rate of
the Gamow-Teller component in Tritium~\cite{CarlsRSW2}.  The result was
that the exchange current increased the cross section of the reaction
$^1{\rm H}(p,e^+\, \nu_e)^2{\rm H}$ by 1.5 \% and that of the reaction
$^3{\rm He}(p,e^+\, \nu_e)^4{\rm He}$ by almost a factor 5. 
The axial exchange current contributions to the solar burning reaction
$^1{\rm H}(p,e^+\, \nu_e)^2{\rm H}$ have later been recalculated on the basis
of $\chi$EFT, and have been found to enhance the cross section obtained
with single-nucleon currents by $\sim 4$ \%~\cite{PaKuMinRho}---however,
see below for a more recent assessment.
Finally, parameter free calculations of the cross section for this reaction
and the associated $^3{\rm He}(p,e^+\, \nu_e)^4{\rm He}$ 
reaction have been carried out in Ref.~\cite{Parketal}.

\subsection{Nuclear charge form factors}

The phenomenological success of the exchange current operators described
above was mainly due to the chiral Lagrangians for the pion-nucleon couplings.
The corresponding contributions to nuclear charge operators involve terms
of higher power in the momentum transfer. The observation by Kloet and Tjon
that there is a significant pion exchange contribution to the charge form factors
of the trinucleons therefore came as a surprise~\cite{KloetTj}.  This pion exchange
operator brings the first diffraction minimum to lower values of momentum transfer
and therefore closer to the experimental data.  This observation was subsequently
confirmed by a calculation of the charge form factor of the $\alpha$-particle~\cite{BorysR}.
Later calculations of the charge form factors of $^3$H, $^3$He, and $^4$He
with realistic variational wave functions reaffirm the need for a substantial exchange
charge effect for agreement with the empirical values~\cite{SPAndR}.  This exchange
current effect is most prominent in the lightest nuclei, and less so in the case of heavier
nuclei as $^{16}$O and $^{40}$Ca, where the shell structure is most prominent~\cite{RadRis,Lovato13}.
The pion exchange effect nevertheless does improve slightly the agreement between
the calculated and empirically extracted charge distributions throughout the periodic
table~\cite{NegRis}.  The best indication of the role of the pion exchange charge
operator was finally provided by the measurement of the charge form factor of the
deuteron~\cite{JLABt20}. Inclusion of the exchange charge operator clearly improved
the agreement with the experimental values.  While the exchange charge operators
that involve two nucleons were found to give substantial contributions to nuclear
charge form factors, the corresponding exchange charge operators that involve
three nucleons were found to give only minor contributions, because of cancellations
between the pion and rho-meson exchange operators. Those involving 4 nucleons were
found to be insignificant~\cite{RadRis2}.

\subsection{The axial charge operator}

That there is a significant nuclear enhancement  of the axial charge of the nucleon was demonstrated by
Kirchbach, Mach, and Riska, who extended the Adler-Weisberger sum rule to light nuclei~\cite{MarMDor}. 
A subsequent explicit calculation of the nuclear enhancement of the axial charge based on meson exchange
indicated that pion-exchange mechanisms yield enhancements of the order 45--60\%. In combination
with short range mechanisms the total enhancement in heavy nuclei is of the order 85--100\%, depending
on the interaction model~\cite{MarRisTsu,Town}. This is sufficient to explain almost all of the empirically
observed $\sim 100$\% enhancement of first forbidden $\beta$-decay transitions in the lead region~\cite{EWar}.
The large nuclear enhancement of the axial charge operator has later been explained within
$\chi$EFT, which provides a dynamical basis for the utility of early chiral charge algebra~\cite{ParkMinRho2}.
Finally, Lee and Riska employed the PCAC relation between the axial current and pion-production
operators to show that the axial exchange current could explain the large difference between
the empirical cross section for the reaction $^1{\rm H}(p,\pi^0)\,pp$ and the value given by the
single-nucleon pion-production operators alone~\cite{LeeRis}.

\subsection{Gerry Brown and the Skyrmion}

In the late 1970's Gerry Brown engaged in a project to develop a chiral dynamics basis for
quark bag models of baryons~\cite{BroRho79}.  In the limit of a small bag radius this led
to a connection to SkyrmeÕs topological soliton model for the nucleon~\cite{BroJaRhoV},
which had been shown by Witten to give results that are consistent with quantum
chromodynamics (QCD) in the large
color limit~\cite{Witt79}.  Skyrme's topological soliton model is based on a chiral Lagrangian
for pions yielding finite size solutions, which may be interpreted as baryons.  The Noether
currents of this Lagrangian can be viewed as models for the electromagnetic and axial currents
of  nucleons and nuclei.  When the meson field is described by the common product ans\"atz for
the two-nucleon system, the current operators separate into single nucleon and exchange
current operators.  Indeed, Nyman and Riska~\cite{NymRis,NymRis2} showed
that if the chiral profile of the Skyrmion field for a nucleon is fitted to its electric form factor, then
the deuteron form factors can be calculated with good qualitative agreement with data.
In the case of the magnetic form factor the exchange current contribution
is large, and essential for agreement with the empirical form factor at large values of
moment transfer.  In the long-range limit there is a simple relation between the isoscalar exchange
current and the phenomenological exchange current that is associated with the $\rho\pi\gamma$
vertex~\cite{WakWei}.  The long-range component of the isovector magnetic moment operator
too is similar to the corresponding conventional pion-exchange magnetic-moment
operator~\cite{BlomRis}.

\section{The chiral effective field theory approach}
\label{sec:theory}

The last two decades have witnessed remarkable developments
in nuclear $\chi$EFT, originally proposed by
Weinberg~\cite{Weinberg90_a,Weinberg90_b,Weinberg90_c}. 
Chiral symmetry is an approximate symmetry of QCD, which
becomes exact in the limit of vanishing quark masses.  Nuclear $\chi$EFT
is the theoretical framework that permits
the derivation of nuclear interactions and electroweak currents with hadronic
degrees of freedom, while preserving the symmetries of QCD---the exact
Lorentz, parity, and time-reversal symmetries, and the
approximate chiral symmetry. The latter requires the pion 
couplings to hadrons to be proportional to powers of its momentum
$Q$ and, as a consequence, the Lagrangian for these interactions may
be expanded in powers of $Q/\Lambda_\chi$, where $\Lambda_\chi\sim 1$ GeV
is the chiral symmetry breaking scale. The Lagrangians may be ordered into
classes defined by the power of $Q/\Lambda_\chi$, or equivalently by order
of the gradients of the pion field and/or pion mass factors. Each of these
contain a certain number of parameters or ``low-energy constants'' (LECs), which
in practice are fixed by comparison with experimental data. These LECs
could in principle be calculated from the underlying QCD theory of quarks
and gluons, but the non-perturbative nature of the theory at low energies
makes this task extremely difficult. Thus, nuclear $\chi$EFT provides a direct connection
between QCD and the strong and electroweak interactions in nuclei, and at the same
time a practical calculational scheme which, at least in principle, may be improved
systematically.  In this sense it provides a fundamental basis for low-energy nuclear physics.

The nuclear $\chi$EFT approach has been applied in a number of studies to derive the two- and three-nucleon potentials\cite{Ord95,Epelbaum98,Ent03,Machleidt11,Nav07,Epe02,Bira94,Bern11,Gir11}
including isospin-symmetry-breaking corrections~\cite{Fri99,Epe99,Fri04,Fri05}.
In the electroweak sector there have been derivations of parity-violating two-nucleon potentials
induced by hadronic weak interactions~\cite{Hax13,Zhu05,Gir08,Viv14}, constructions of
nuclear electroweak currents~\cite{ParkMinRho2}, and studies of Compton scattering on
nucleons and nuclei with the explicit inclusion of $\Delta$-resonance degrees of
freedom~\cite{Pascalutsa2003,Griesshammer2012}.
Here the focus is on nuclear electroweak current operators. These were originally
derived up to one loop level in the heavy-baryon formulation of covariant perturbation
theory by Park {\it et al.}~\cite{ParkMinRho2,Park96,Parketal}.  More recently two
independent derivations, based on time-ordered perturbation theory (TOPT),
have been published---one by the present authors (RS)~\cite{Pastore09,Pastore11,Piarulli13,Baroni16}
and the other by K\"olling {\it et al.}~\cite{Koelling09,Koelling11}, although these latter works only deal with
electromagnetic currents.  In the following, we outline the derivation of these electroweak
operators, referring to the
original papers~\cite{Pastore09,Pastore11,Piarulli13,Baroni16} for the more technical aspects.   

\subsubsection{Interaction Hamiltonians}
\label{sec:hint}

In the simplest implementation, $\chi$EFT Lagrangians are constructed
in terms of nucleon and pion degrees of freedom. This has been described
in a number of papers~\cite{Gasser84,Fettes00}, and $\pi \pi$  and $\pi N$ Lagrangians,
denoted respectively as ${\cal L}^{(m)}_{\pi \pi}$ and ${\cal L}^{(n)}_{\pi N}$,
have been derived up to high order in the chiral expansion. Contributions
that arise from additional degrees of freedom, such as
$\Delta$-resonances and heavier mesons, are subsumed in the
LECs of ${\cal L}^{(n)}_{\pi N}$ and ${\cal L}_{\pi\pi}^{(m)}$.
In principle these Lagrangians contain an infinite number of interactions
compatible with the QCD symmetries, but as the transition
amplitudes obtained from them may be expanded in powers of $Q/\Lambda_\chi$,    
the number of terms that contribute to the
amplitude at any given order of the expansion is finite~\cite{Weinberg90_a,Weinberg90_b,Weinberg90_c}.
The Hamiltonians are constructed  from the chiral Lagrangians by
the canonical formalism.

The leading interaction terms in
${\cal L}^{(1)}_{\pi N}$, ${\cal L}^{(2)}_{\pi N}$, and ${\cal L}^{(3)}_{\pi N}$ in the $\pi N$
sector, and ${\cal L}^{(2)}_{\pi \pi}$ and ${\cal L}^{(4)}_{\pi \pi}$ in the $\pi\pi$ sector, which are relevant to the
derivation of nuclear potentials and electroweak operators at one loop level lead
to the following Hamiltonians:
\begin{eqnarray}
\!\!\!\!\!\!\!\!&&H_{\pi N} =\int\! {\rm d}{\bf x}\,
 N^\dagger\Big[ \frac{g_A}{2 f_\pi}\,\tau_a\,
 {\bm \sigma} \cdot {\bm \nabla} \pi_a \!+\!\frac{1}{4f_\pi^2}
{\bm \tau}\cdot
({\bm \pi}\times  {\bm \Pi})+\dots \Big]N \ ,
\label{eq:hpiN} \\
\!\!\!\!\!\!\!\!&&H_{\gamma N}= e\int\!{\rm d}{\bf x}\, N^\dagger \Big[ e_N \, V^0 +
i \,\frac{e_N}{2\,m} 
\left( -\overleftarrow{\bm \nabla}\cdot {\bf V}
+ {\bf V} \cdot \overrightarrow{\bm \nabla} \right) 
-\frac{\mu_N}{2\, m} \, {\bm \sigma}\cdot  {\bm \nabla} \times {\bf V} \nonumber\\
\!\!\!\!\!\!\!\!&&\qquad - \frac{2\, \mu_N - e_N }{8\, m^2} 
\Big( {\bm \nabla }^2 V^0  + {\bm \sigma} \times  {\bm \nabla} V^0 \cdot \overrightarrow{\bm \nabla}
- \overleftarrow{\bm \nabla} \cdot {\bm \sigma} \times  {\bm \nabla} V^0 \Big) +\dots \Big] N \ ,
\label{eq:hgn}  \\
\!\!\!\!\!\!\!\!&&H_{\gamma \pi} =e\int\!{\rm d}{\bf x}\,\left[
V^0\left( {\bm \pi} \times  {\bm \Pi} \right)_z +\epsilon_{zab}\,
 \pi_a\, \left( {\bm \nabla} \pi_b\right) \cdot  {\bf V}+
\dots \right] 
\label{eq:hgpi}\ , \\
\!\!\!\!\!\!\!\!&&H_{ \gamma \pi N} = \frac{e}{2f_\pi} \int\!{\rm d}{\bf x}\, N^\dagger\Big[ 
  \frac{ g_A}{2\, m} \left( {\bm \tau}\cdot {\bm \pi}+\pi_z \right)\,
    {\bm \sigma}\cdot {\bm \nabla} V^0-\big[   8\,d_8 {\bm \nabla} \pi_z \nonumber\\
\!\!\!\!\!\!\!\! &&\qquad + 8\, d_9\,  \tau_a  {\bm \nabla} \pi_a 
 -\left( 2\, d_{21}-d_{22} \right) \, \epsilon_{zab} \tau_b\, {\bm \sigma}\times
 {\bm \nabla}\pi_a  \big]  \cdot {\bm \nabla} \times {\bm V} \,
+\dots  \Big] N \ ,
\label{eq:hgpin} \\
\!\!\!\!\!\!\!\!&&H_{ A N}=\frac{g_A}{2} \int {\rm d}{\bf x}\, N^\dagger
\left(  \tau_a \, {\bm \sigma} \cdot {\bf A}_a  +\dots \right) N \ ,\\
\!\!\!\!\!\!\!\!&&H_{ A \pi}=f_\pi \int
{\rm d}{\bf x} \left(  {\bf A}_a \cdot {\bm \nabla}\pi_a+ A^0_a  \, \Pi_a + \dots \right) \ ,\\
\!\!\!\!\!\!\!\!&&H_{ A \pi N}=\int {\rm d}{\bf x}\, N^\dagger
\Big[ -\frac{1}{4f_\pi}A^0_a \left({\bm \tau}\times {\bm \pi}\right)_a
+\frac{2\,c_3}{f_\pi}{\bf A}_a \cdot {\bm \nabla} \pi_a \nonumber\\
\!\!\!\!\!\!\!\!&&\,\,\,\,\,\,\,\, -\frac{c_4}{f_\pi}\epsilon_{abc} \tau_a ({\bf A}_b\times {\bm \nabla}\pi_c) 
\cdot {\bm \sigma}  -\frac{c_6}{4mf_\pi}({\bm \tau}\times{\bm \pi})_a 
\left({\bm \nabla} \times {\bf A}_a\right)\cdot{\bm \sigma}+ \dots \Big]N\ ,
 \end{eqnarray}
where $g_A $, $f_\pi$, $e$, and $m$ are, respectively, the
nucleon axial coupling constant, pion decay amplitude, proton electric charge, and
nucleon mass, and the parameters $c_i$ and $d_i$ are LECs in 
the ${\cal L}_{\pi N}^{(2)}$ and ${\cal L}_{\pi N}^{(3)}$ Lagrangians.

The isospin doublet of (non-relativistic) nucleon fields, isospin triplet of pion fields and conjugate
fields, electromagnetic vector field and weak axial field are
denoted by $N$, ${\bm \pi}$ and ${\bm \Pi}$, $V^\mu$, and $A^\mu$ respectively, and
${\bm \sigma}$ and ${\bm \tau}$ are spin and isospin Pauli
matrices.  The arrow over the gradient specifies whether it acts on
the left or right nucleon field.  The isospin operators $e_N$ and $\mu_N$
are defined as
\begin{equation}
e_N = (1+\tau_z)/2 \ , \,\,\,
\kappa_N =  (\kappa_S+ \kappa_V \, \tau_z)/2 \ , \,\,\, \mu_N = e_N+\kappa_N  \ ,
\label{eq:ekm}
\end{equation}
where $\kappa_S$ and $\kappa_V$ are the isoscalar and isovector combinations
of the anomalous magnetic moments of the proton and neutron.
The power counting of the resulting vertices follows by noting that each
gradient brings in a factor of $Q$, so, for example, the two terms in $H_{\pi N}$
are both of order $Q$, while (ignoring the counting $Q$ assumed for the
external fields $V^\mu$ and $A^\mu$) the first term in $H_{\gamma \pi N}$ ($H_{A\pi N}$) is
of order $Q$ ($Q^0$), while the remaining ones
are of order $Q^2$ ($Q^1$).

In addition to the chiral Hamiltonians above, up to and including order
$Q^2$ there are fourteen contact interaction terms allowed by the
symmetries of the strong interactions, each one multiplied by a LEC. Two of these contact
terms (proportional to the LECs $C_S$ and $C_T$ in standard notation) are of a
non-derivative type, and therefore are of order $Q^0$,
while the remaining twelve (proportional to the LECs $C_i^\prime$) of order $Q^2$
involve two gradients acting on the nucleon fields (these are listed in Ref.~\cite{Gir10}).
The contact potential at order $Q^2$, derived from them
in the two-nucleon center-of-mass system, in fact depends on $C_S$ and $C_T$,
and seven linear combinations of the $C_i^\prime$, which are customarily denoted
as $C_1,\dots,C_7$. The remaining five linear combinations of $C_i^\prime$
have been shown to be related to $C_S$ and $C_T$ by requiring
that the Poincar\'e covariance of the theory be satisfied to order
$Q^2$~\cite{Gir10}. The $Q^2$
potential therefore involves nine independent LECs. (As a side remark, the contact
potential at order $Q^4$ requires an additional fifteen independent LECs.)
These LECs are determined by fits to two-nucleon elastic scattering data.

Minimal substitution in the gradient terms leads to a (two-nucleon)
electromagnetic contact current denoted as
${\bf j}^{(1)}_{\gamma,\rm min}$ in Refs.~\cite{Pastore09,Piarulli13},
where the superscript ${(n)}$ specifies the power counting $Q^n$.
Non-minimal couplings through the electromagnetic field tensor
$F_{\mu \nu}$ are also allowed.  It may be shown~\cite{Pastore09} that only
two independent operator structures enter at order $Q^1$, 
which lead to the contact term given by
\begin{eqnarray}
{\bf j}^{(1)}_{\gamma,\rm nm}&=& - i\,e \Big[ C_{15}^\prime\, {\bm \sigma}_1 
+C_{16}^\prime \times (\tau_{1,z} - \tau_{2,z})\,{\bm \sigma}_1  \Big]\times {\bf q}  
+ \left(1 \rightleftharpoons 2\right)\,
\label{eq:gnm}
\end{eqnarray}
where ${\bf q}$ is the external field momentum, and the isoscalar $C_{15}^\prime$
and isovector $C_{16}^\prime$ LECs (as well as the $d_i$'s multiplying the higher
order terms in the $\gamma\pi N$ Hamiltonian) can be determined by fitting
photo-nuclear data in the few-nucleon systems~\cite{Piarulli13}.

In the weak axial sector, there is a single contact term at order $Q^0$,
\begin{equation}
\label{eq:jctct}
{\bf j}_{5,a}^{(0)}=z_0\left({\bm \tau}_1\times{\bm \tau}_2\right)_a \left[{\bm \sigma}_1\times{\bm \sigma}_2-\frac{{\bf q}}{q^2+m_\pi^2}\, {\bf q}\cdot\left({\bm \sigma}_1\times{\bm \sigma}_2\right) \right]\,
\end{equation}
(here the second term of Eq.~(\ref{eq:jctct}) is the pion-pole contribution),
and none at order $Q^1$. This term is due to an interaction of the type
$\left(\overline{N}\gamma^\mu\gamma_5\,  N\right) \left(\overline{N} u_\mu\,N\right) $ and,
as first pointed out in Ref.~\cite{Gar06}, the LEC $z_0$ is related to
the LEC $c_D$ (in standard notation), which enters the three-nucleon potential at
leading order.  The two LECs $c_D$ and $c_E$ which fully characterize this
potential have recently been constrained by reproducing the empirical value
of the Gamow-Teller matrix element in tritium $\beta$ decay and the binding
energies of the trinucleons~\cite{Gazit09,Marcucci12} (see below).  Lastly, in the
limit of small momentum transfers,
there are two independent two-nucleon contact terms in the axial
charge at order $Q^1$~\cite{Baroni16}:
\begin{eqnarray}
\label{eq:4.22}
\rho^{(1)}_{5,a} &=&i\, z_1 \left({\bm\tau}_1\times{\bm\tau}_2\right)_a\,
\left({\bm \sigma}_1\cdot{\bf k}_1-{\bm \sigma}_2\cdot{\bf k}_2\right) \nonumber  \\
&&+i\,z_2\left({\bm \sigma}_1\times{\bm \sigma}_2\right)\cdot\left(\tau_{1,a}\, {\bf k}_2-\tau_{2,a}\, {\bf k}_1\right)\ .
\end{eqnarray}
The LECs $z_1$ and $z_2$ have, however yet to be determined.

\subsubsection{From amplitudes to potentials and currents}
\label{sec:road}
Application of $\chi$EFT to nuclear structure and bound states requires going beyond perturbation theory. As suggested by 
Weinberg ~ \cite{Weinberg90_a,Weinberg90_b,Weinberg90_c},
the formalism briefly described below for constructing nuclear potentials and currents is based on
time-ordered perturbation theory (TOPT) although it differs from Weinberg's
in the way reducible contributions are dealt with. This has been employed in
Refs.~\cite{Pastore08,Pastore09,Pastore11,Piarulli13,Baroni16} for constructing nuclear potentials and currents. 

The terms in the TOPT expansion are conveniently represented by diagrams.
Here a distinction is made between reducible diagrams, which involve
at least one pure nucleonic intermediate state, and irreducible diagrams,
which include pionic and nucleonic intermediate states. The contributions of the former are enhanced with respect to those of corresponding irreducible ones
by a factor of $Q$ for each pure nucleonic intermediate state.  In the static limit,
in which $m \rightarrow \infty$ or, equivalently, nucleon kinetic
energy terms are dropped, the reducible contributions are infrared-divergent.  The prescription proposed by Weinberg~\cite{Weinberg90_a,Weinberg90_b,Weinberg90_c} to
treat these is to define the nuclear potential (and currents) as given by the
irreducible contributions only. The reducible contributions are generated
by solution of the Lippmann-Schwinger (or Schr\"odinger) equation iteratively, with
the nuclear potential (and currents) given by the irreducible amplitudes.

The formalism originally developed in Ref.~\cite{Pastore08} is based on this approach. The omission of the reducible contributions from the definition of the interaction operators requires care, when the irreducible amplitudes are evaluated in the static approximation, which is commonly used. The iterative process will in that limit generate only part of the reducible amplitude. The reducible part of the amplitude beyond the static approximation has then to be incorporated order by order---along with the irreducible amplitude---in
the definition of nuclear operators.  This scheme in combination with TOPT, which is best suited to separate the reducible content from the irreducible one, has been implemented
in Refs.~\cite{Pastore09,Pastore11,Piarulli13,Baroni16} and is described below.  The method does however lead to nuclear operators, which are not uniquely defined because of the non-uniqueness of the
transition amplitude off-the-energy shell. This lack of uniqueness is immaterial, however, because the resulting operators are
unitarily equivalent, and therefore the description of physical observables is not affected by this
ambiguity ~\cite{Pastore11}. 

Another approach for overcoming the difficulties posed by the reducible amplitudes,
has been introduced by Epelbaum and collaborators~\cite{Epelbaum98}. That method
is usually referred to as the unitary transformation method and is based on TOPT.  It exploits the Okubo
(unitary) transformation~\cite{Okubo54} to decouple the Fock space of pions and nucleons
into two subspaces, one that has pure
nucleonic states and the other with
states which retain at least one pion. In this decoupled space, the amplitude
does not involve enhanced contributions associated with the reducible diagrams.
The subspaces are not uniquely defined, as it is always possible to perform
additional unitary transformations on them, with a consequent change in the formal definition
of the resulting nuclear operators. This, of course, does not affect the calculated physical observables.

The two TOPT-based methods outlined above lead to formally equivalent operator structures for the nuclear potential and electromagnetic currents up to loop-corrections included~\cite{Piarulli13}. It is natural to conjecture that the two methods are closely related, although this remains to be proved. Below we briefly outline the methods developed
in Refs.~\cite{Pastore09,Pastore11,Piarulli13,Baroni16} and sketch how nuclear operators are
derived from transition amplitudes.

We start from the conventional perturbative expansion of the
$NN$ scattering amplitude $T$:
\begin{equation}
 \langle f \!\mid T\mid\! i \rangle= 
 \langle f\! \mid H_1 \sum_{n=1}^\infty \left( 
 \frac{1}{E_i -H_0 +i\, \eta } H_1 \right)^{n-1} \mid\! i \rangle \ .
\label{eq:pt}
\end{equation}
Here $\mid\! i \rangle$ and $\mid\!\! f \rangle$ represent the initial
and final $NN$ states with energy $E_i=E_f$, $H_0$ is the Hamiltonian
describing free pions and nucleons, and $H_1$ is the Hamiltonian
describing interactions between them (see Sec.~\ref{sec:hint}).
The evaluation of this amplitude is in practice carried out by inserting
complete sets of $H_0$ eigenstates between successive terms of $H_1$.
Power counting is then used to organize the expansion.

In the perturbation expansion of Eq.~(\ref{eq:pt}), a generic (reducible or
irreducible) contribution is characterized by a certain number, say
$M$, of vertices, each scaling as $Q^{\alpha_i}\times Q^{-\beta_i/2}$
($i$=$1,\dots,M$), where $\alpha_i$ is the power counting implied by the
relevant interaction Hamiltonian and $\beta_i$ is the number of
pions in and/or out of the vertex, a corresponding $M-1$ number of energy
denominators, and possibly $L$ loops. Out of these $M-1$ energy
denominators, $M_K$ will involve only nucleon kinetic
energies, which scale as $Q^2$, and the remaining $M-M_K-1$ will involve,
in addition, pion energies, which are of order $Q$.  Loops, on the other hand,
contribute a factor $Q^3$ each, since they imply integrations over intermediate
three momenta. Hence the power counting associated with such a contribution is
\begin{equation}
\left(\prod_{i=1}^M  Q^{\alpha_i-\beta_i/2}
\right)\times \left[ Q^{-(M-M_K-1)}\, Q^{-2M_K} \right ]
\times Q^{3L} \ .
\label{eq:count}
\end{equation}
Clearly, each of the $M-M_K-1$ energy denominators can be further expanded as
\begin{equation}
\frac{1}{E_i-E_I-\omega_\pi}= -\frac{1}{\omega_\pi}
\bigg[ 1 + \frac{E_i-E_I}{\omega_\pi}+
\frac{(E_i-E_I)^2}{\omega^2_\pi} + \dots\bigg] \ ,
\label{eq:deno}
\end{equation}
where $E_I$ denotes the kinetic energy of the intermediate two-nucleon state,
$\omega_\pi$ the pion energy (or energies, as the case may be), and the
ratio $(E_i-E_I)/\omega_\pi$ is of order $Q$.  The terms proportional to
powers of $(E_i-E_I)/\omega_\pi$ lead to non-static corrections.

The $Q$-scaling of the interaction vertices and the considerations 
above show that the amplitude $T$ admits the following expansion:
\begin{equation}
 T=T^{(\nu)} + T^{(\nu+1)} + T^{(\nu+2)} + \dots \ ,
\label{eq:tmae}
\end{equation}
where $T^{(n)} \sim Q^n$, and chiral symmetry ensures that
$\nu$ is finite. In the case of the two-nucleon potential 
$\nu=0$.  A two-nucleon potential $v$ can then be derived,
which when iterated in the Lippmann-Schwinger (LS) equation,
\begin{equation}
v+v\, G_0\, v+v\, G_0 \, v\, G_0 \, v +\dots \ ,
\label{eq:lse}
\end{equation}
leads to the on-the-energy-shell ($E_i=E_f$) $T$-matrix in Eq.~(\ref{eq:tmae}),
order by order in the power counting.  In practice, this requirement can only be
satisfied up to a given order $n^*$, and the resulting potential, when inserted into
the LS equation, will generate contributions of order $n > n^*$, which do not match $T^{(n)}$.
In Eq.~(\ref{eq:lse}), $G_0$ denotes the free two-nucleon propagator, $G_0=1/(E_i-E_I+i\, \eta)$,
and we assume that
\begin{equation}
v=v^{(0)}+v^{(1)}+v^{(2)}+\dots\ ,
\end{equation}
where the still to be determined term $v^{(n)}$ is of order $Q^n$.  We also note that, generally, a term
like $v^{(m)}\, G_0 \, v^{(n)}$ is of order $Q^{m+n+1}$, since $G_0$ is of order $Q^{-2}$
and the implicit loop integration brings in a factor $Q^3$. 

Having established the above power counting,
we obtain
\begin{eqnarray}
v^{(0)} &=& T^{(0)} \ , \label{eq:v0}\\
v^{(1)} &=& T^{(1)}-\left[ v^{(0)}\, G_0\, v^{(0)}\right] \ , \\
v^{(2)} &=& T^{(2)}-\left[ v^{(0)}\, G_0\, v^{(0)}\, G_0\, v^{(0)}\right] \nonumber\\
&&\qquad-\left[ v^{(1)}\, G_0 \, v^{(0)}
+v^{(0)}\, G_0\, v^{(1)}\right] \ . \label{eq:v2}
\end{eqnarray}
The leading-order (LO) $Q^0$ term, $v^{(0)}$, consists of (static) 
one-pion exchange (OPE) and two (non-derivative) contact interactions, 
while the next-to-leading (NLO)
$Q^1$ term, $v^{(1)}$, is easily seen to vanish~\cite{Pastore11}, since the leading non-static corrections
$T^{(1)}$ to the (static) OPE amplitude add up to zero on the energy
shell, while the remaining diagrams in $T^{(1)}$ represent iterations of $v^{(0)}$,
whose contributions are exactly canceled by $\left[ v^{(0)}\, G_0\, v^{(0)}\right]$
(complete or partial cancellations of this type persist at higher $n\ge 2$ orders).
The next-to-next-to-leading (N2LO) $Q^2$ term, which follows from Eq.~(\ref{eq:v2}),
contains two-pion-exchange (TPE) and contact (involving two gradients of the nucleon fields)
interactions. 

The inclusion (in first order) of electroweak interactions in the perturbative
expansion of Eq.~(\ref{eq:pt}) is in principle straightforward.
The transition operator can be expanded as~\cite{Pastore11,Baroni16}:
\begin{equation}
T_{\rm ext}=T_{\rm ext}^{(\nu_e)}+T_{\rm ext}^{(\nu_e+1)}+T_{\rm ext}^{(\nu_e+2)} +\dots \ ,
\end{equation}
where $T_{\rm ext}^{(n)}$ is of order $Q^n$ and $\nu_e=-3$ in this case.
The nuclear electromagnetic (weak axial) charge, $\rho_\gamma$ ($\rho_{5,a}$),
and current, ${\bf j}_\gamma$ (${\bf j}_{5,a}$), operators follow
from $v_\gamma= V^0\, \rho_\gamma-{\bf V}\cdot {\bf j}_\gamma$
($v_5= A^0_a\, \rho_{5,a}-{\bf A}_a\cdot {\bf j}_{5,a}$), where
$V^\mu=(V^0,{\bf V})$ [$A_a^\mu=(A_a^0,{\bf A}_a)$] is the electromagnetic vector (weak axial)
field, and it is assumed that $v_{\rm ext}$ has a similar expansion
as $T_{\rm ext}$.  The requirement that, in the context of the LS equation,
$v_{\rm ext}$ matches $T_{\rm ext}$ order by order in the power counting
implies relations for the $v^{(n)}_\gamma=V^0\, \rho_\gamma^{(n)}-{\bf V}\cdot {\bf j}_\gamma^{(n)}$
and  $v^{(n)}_5=A^0_a\, \rho_{5,a}^{(n)}-{\bf A}_a\cdot {\bf j}_{5,a}^{(n)}$,
which can be found in Refs.~\cite{Pastore11,Baroni16}, similar to those derived above
for $v^{(n)}$, the strong-interaction potential.  

The lowest order terms that contribute to the electromagnetic charge and axial current
operators have $\nu_e=-3$,
\begin{eqnarray}
\label{eq:r-3}
\rho_\gamma^{(-3)}&=&e\, \frac{ 1+\tau_{1,z}}{2} + (1 \rightleftharpoons 2)  , \\
{\bf j}_{5,a}^{(-3)}&=& -\frac{g_A}{2} \, \tau_{1,a}
\left( {\bm\sigma}_1-\frac{{\bf q}}{q^2+m_\pi^2}\,{\bm\sigma}_1\cdot{\bf q} \right)
+ \left( 1 \rightleftharpoons 2\right) \ .
\label{eq:r-3a}
\end{eqnarray}
There are no $Q^{-3}$ contributions to ${\bf j}$ and $\rho_{5,a}$, and the lowest order ($\nu_e=-2$) consists
of electromagnetic current and axial charge operators, given by
\begin{eqnarray}
{\bf j}_\gamma^{(-2)}&=&\frac{e}{2\, m}
\left(2\,  {\bf K}_1 \, \frac{ 1+\tau_{1,z}}{2} +i\,{\bm \sigma}_1\times {\bf q }\, 
 \frac{ \mu^S+\mu^V \tau_{1,z}}{2} \right)+
 (1 \rightleftharpoons 2)\ , \label{eq:jlo} \\
 \rho_{5,a}^{(-2)}&=& -\frac{g_A}{2\,m} \, \tau_{1,a}\, {\bm\sigma}_1
\cdot {\bf K}_1
+ \left(1 \rightleftharpoons 2\right) \ ,
\label{eq:rr-2} 
\end{eqnarray}
where
${\bf k}_i$ and ${\bf K}_i$ denote hereafter the combinations of
initial and final nucleon momenta
\begin{equation}
{\bf k}_i={\bf p}_i^\prime-{\bf p}_i \ ,\qquad {\bf K}_i=({\bf p}_i^\prime+{\bf p}_i)/2\ .
\end{equation}
The counting $Q^{-3}$ ($Q^{-2}$) in the electromagnetic charge and axial
current (electromagnetic current and axial
charge) operators follows from the product of the power counting
associated with the $\gamma NN$, $A NN$, $A\pi$, and $\pi NN$ vertices,
and the $Q^{-3}$ factor due to the momentum-conserving $\delta$-function implicit
in disconnected terms of this type. 

\begin{figure}[t]
\vspace*{0.6cm}
\centerline{\includegraphics[width=11cm]{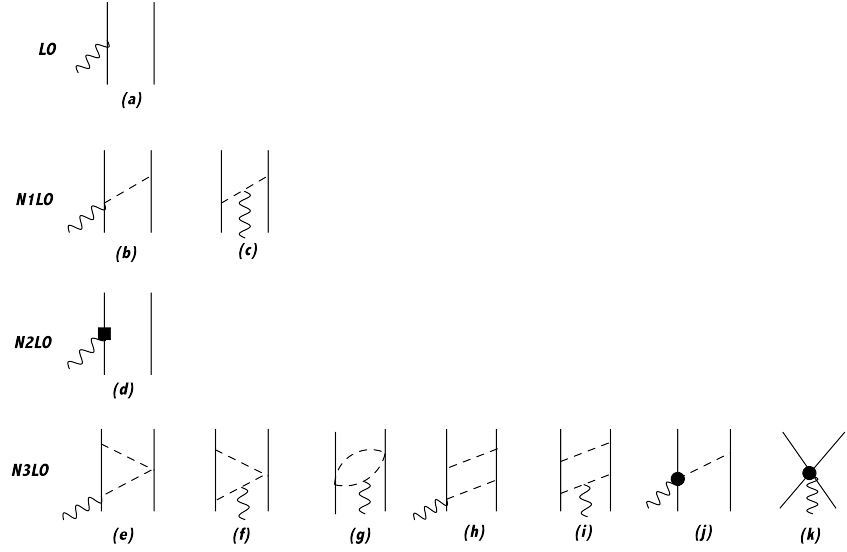}}
\vspace{8pt}
\caption{Diagrams illustrating one- and two-body electromagnetic currents entering at $Q^{-2}$ (LO),
$Q^{-1}$ (NLO), $Q^{\,0}$ (N2LO), and $Q^{\,1}$ (N3LO).  Nucleons, pions,
and photons are denoted by solid, dashed, and wavy lines, respectively.  The square
in panel (d) represents the $(Q/m)^2$ relativistic correction to the LO one-body current;
the solid circle in panel (j) is associated with the $\gamma \pi N$ current coupling
of order $Q$, involving the LECs $d_8$, $d_9$, and $2\,d_{21}-d_{22}$;
the solid circle in panel (k) denotes two-body contact terms of minimal and non-minimal
nature, the latter involving the LECs $C_{15}^\prime$ and $C_{16}^\prime$.
Only one among all possible time orderings is shown for the NLO and N3LO currents,
so that all direct- and crossed-box contributions are accounted for.}
\label{fig:f2}
\end{figure}

\begin{figure}[t]
\vspace*{0.6cm}
\centerline{\includegraphics[width=9cm]{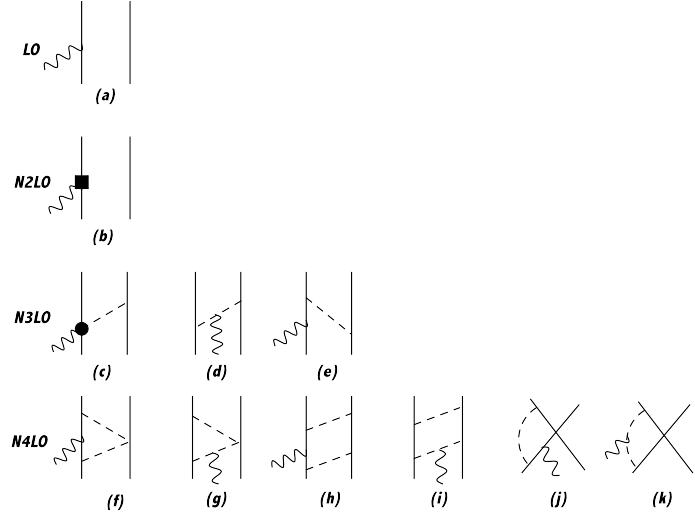}}
\vspace{8pt}
\caption{Diagrams illustrating one- and two-body electromagnetic charge operators
entering at $Q^{-3}$ (LO), $Q^{-1}$ (N2LO), $Q^{0}$ (N3LO), $Q^{1}$ (N4LO).
The square in panel (b) represents the $(Q/m)^2$ relativistic correction to the LO
one-body charge operator, whereas the solid circle in panel (c) is associated with
a $\gamma \pi N$ charge coupling of order $Q$.
As in Fig.~\ref{fig:f2}, only a single time ordering is shown for the N3LO and N4LO charge operators.}
\label{fig:f5}
\end{figure}

\begin{figure}[t]
\vspace*{0.6cm}
\centerline{\includegraphics[width=11cm]{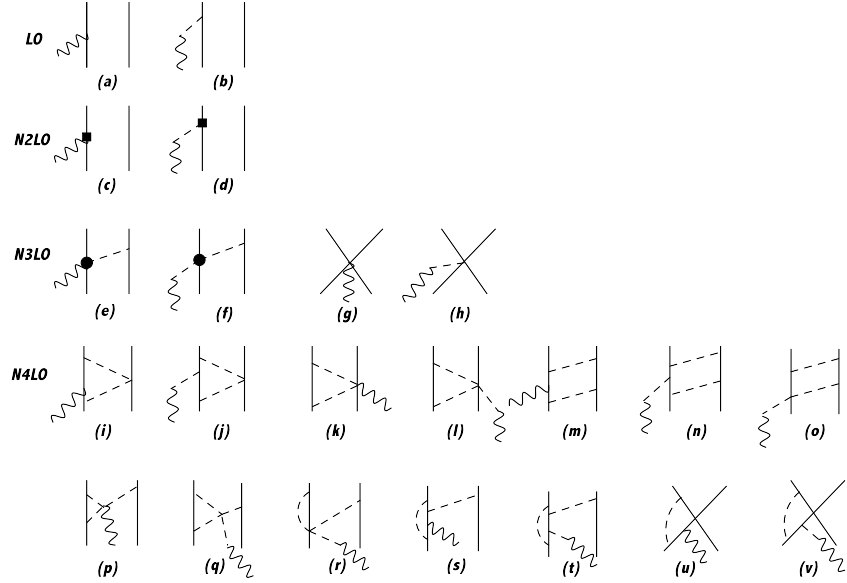}}
\vspace{8pt}
\caption{Diagrams illustrating one- and two-body axial currents entering at $Q^{-3}$ (LO),
$Q^{-1}$ (N2LO), $Q^{\,0}$ (N3LO), and $Q^{\,1}$ (N4LO).  Nucleons, pions,
and axial fields are denoted by solid, dashed, and wavy lines, respectively.
The squares in panels (c) and (d) denote relativistic corrections to the one-body
axial current, while the circles in panels (e) and (f) represent vertices implied
by the ${\cal L}^{(2)}_{\pi N}$ chiral Lagrangian, involving
the LECs $c_i$ (see Ref.~\cite{Baroni16} for additional explanations).
As in Fig.~\ref{fig:f2}, only a single time ordering is shown.}
\label{fig:f2a}
\end{figure}

\begin{figure}[t]
\vspace*{0.6cm}
\centerline{\includegraphics[width=11cm]{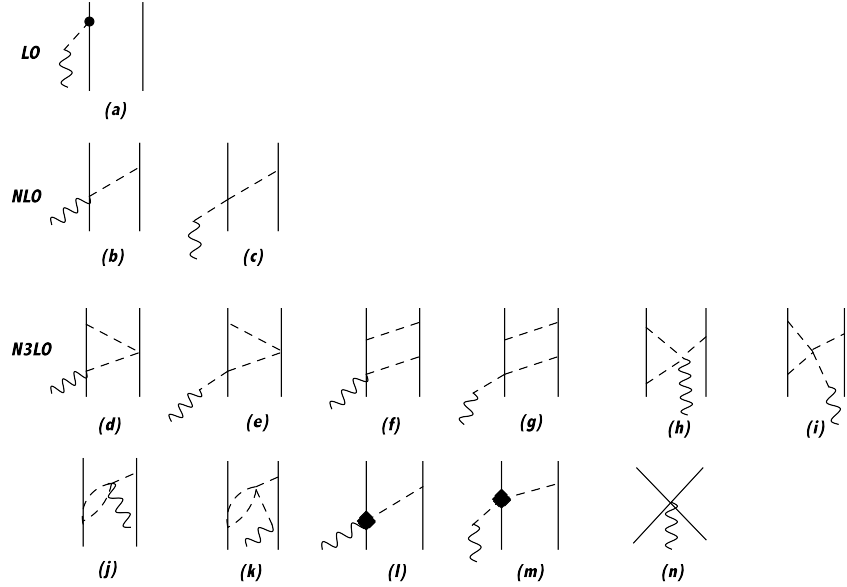}}
\vspace{8pt}
\caption{Diagrams illustrating one- and two-body axial charge operators entering
at $Q^{-2}$ (LO), $Q^{-1}$ (NLO), and $Q^{\,1}$ (N3LO).  Nucleons, pions,
and axial fields are denoted by solid, dashed, and wavy lines, respectively.
The diamonds in panels (l) and (m) indicate higher order $A\pi N$ vertices
implied by the ${\cal L}^{(3)}_{\pi N}$ chiral Lagrangian, involving
the LECs $d_i$ (see Ref.~\cite{Baroni16} for additional explanations).
As in Fig.~\ref{fig:f2}, only a single time ordering is shown.}
\label{fig:f5a}
\end{figure}

The contributions up to one loop to the electromagnetic current and charge operators
are illustrated diagrammatically in Figs.~\ref{fig:f2} and~\ref{fig:f5}, while those to the
weak axial current and charge operators in Figs.~\ref{fig:f2a} and~\ref{fig:f5a}.
As already noted, the LO starts at $\nu_e=-2$ for the electromagnetic current and axial
charge and at $\nu_e=-3$ for the electromagnetic charge and axial current; N$n$LO
corrections to these are labelled as $Q^n \times \,{\rm LO}$.  We begin by discussing the
electromagnetic operators.

The electromagnetic currents from LO, NLO, and N2LO terms and from N3LO loop corrections
depend only on the known parameters $g_A$ and $f_\pi$ (NLO and N3LO), and the nucleon
magnetic moments (LO and N2LO).  Unknown LECs enter the N3LO OPE contribution
involving the $\gamma \pi N$ vertex of order $Q^2$ from $H_{\gamma \pi N}$, the term proportional
to the $d_i$ in Eq.~(\ref{eq:hgpin}), as well as the contact currents implied by non-minimal
couplings, Eq.~(\ref{eq:gnm}),  discussed in the next
subsection.  On the other hand, in the charge operator there are no unknown LECs  up
to one loop level, and OPE contributions, illustrated in panels
(c)-(e) of Fig.~\ref{fig:f5}, only appear at N3LO.
The contributions in panels (d) and (e) involve non-static corrections~\cite{Pastore11}, while the contribution
in panel (c) is associated with the $\gamma \pi N$ coupling of order $Q$ originating
from the first term in Eq.~(\ref{eq:hgpin}).  It leads to a two-body charge operator:
\begin{equation}
\rho^{(0)}_\gamma({\rm OPE}) =\frac{e}{8\, m} \frac{g_A^2}{f_\pi^2} \left( {\bm \tau}_1 \cdot {\bm \tau_2}
+ \tau_{2z}\right)\, \frac{{\bm \sigma}_1 \cdot {\bf q} \,\, {\bm \sigma}_2 \cdot {\bf k}_2}{k^2_2+m_\pi^2}
+ (1 \rightleftharpoons 2) \ .
\label{eq:pich}
\end{equation}
In the present $\chi$EFT context, $\rho^{(0)}_\pi$ was derived first by
Phillips in 2003~\cite{Phillips03}.  However,
the operator of Eq.~(\ref{eq:pich}) 
is the same as the $\pi$-exchange contribution derived within the
conventional approach (see Ref.~\cite{Ris89} and references
therein). This operator plays an important role
in yielding predictions for the $A$=2--4 charge form factors
that are in very good agreement with the experimental data at low 
and moderate values of the momentum transfer 
($q \lesssim 5$ fm$^{-1}$)~\cite{Car98,Piarulli13}.
The calculations in Ref.~\cite{Piarulli13} also showed that
the OPE contributions from panels (d) and (e) of Fig.~\ref{fig:f5}
are typically an order of magnitude smaller than those generated by panel (c).

The axial current and charge operators illustrated in Figs.~\ref{fig:f2a} and~\ref{fig:f5a}
include pion-pole contributions, which are crucial for the current to be conserved
in the chiral limit~\cite{Baroni16} (these contributions were ignored in the earlier studies
of Park {\it et al.}~\cite{ParkMinRho2,Parketal}; obviously, they are suppressed in low momentum
transfer processes). It is also interesting to note that there are no direct couplings
of $A^0_a$ to the nucleon, see panel (a) in Fig.~\ref{fig:f5a}.
In the axial current pion-range contributions
enter at N3LO, panels (e) and (f) of Fig.~\ref{fig:f2a}, and involve vertices
form the sub-leading ${\cal L}^{(2)}_{\pi N}$ Lagrangian, proportional
to the LECs $c_3$, $c_4$, and $c_6$.  It is given by (the complete operator,
including pion pole contributions, is listed in Ref.~\cite{Baroni16})
\begin{eqnarray}
\label{eq:opej1fin}
{\bf j}_{5,a}^{(0)}({\rm OPE})&=& \frac{g_A}{2\,f_\pi^2} \bigg\{
2\, c_3 \, \tau_{2,a}\, {\bf k}_2
+\left({\bm \tau}_1\times{\bm \tau}_2\right)_a 
\bigg[ \frac{i}{2\, m} {\bf K}_1 - \frac{c_6+1}{4\, m}  {\bm \sigma}_1\times{\bf q} \nonumber\\
&&+\left( c_4+\frac{1}{4\, m}\right) {\bm \sigma}_1\times{\bf k}_2 \bigg] 
\bigg\} \frac{{\bm\sigma}_2\cdot{\bf k}_2} {k_2^2+m_\pi^2}+ (1\rightleftharpoons 2)\ .
\end{eqnarray}
In contrast, the axial charge has a OPE contribution at NLO,
illustrated in panels (b) and (c) of Fig.~\ref{fig:f5a}, which reads
\begin{eqnarray}
\label{eq:6.66}
\rho^{(-1)}_{5,a}({\rm OPE})&=&
i\frac{g_A}{4\,f_\pi^2}\left({\bm \tau}_1\times{\bm \tau}_2\right)_a
\frac{{\bm \sigma}_2\cdot{\bf k}_2}{k_2^2+m_\pi^2} + (1\rightleftharpoons 2)\ .
\end{eqnarray}
In fact, an operator of precisely this form was derived by Kubodera
{\it et al.}~\cite{Kubodera78} in the late seventies,
long before the systematic approach based on chiral Lagrangians 
now in use was established.  Corrections to the axial current at N4LO in panels (i)-(v) of
Fig.~\ref{fig:f2a} and those to the axial charge at N3LO in panels (d)-(n) of Fig.~\ref{fig:f5a}
have yet to be included in actual calculations of weak transitions
in nuclei.  It is worthwhile noting that vertices involving three or
four pions, such as those, for example, occurring in panels
(l), (p), (q) and (r) of Fig.~\ref{fig:f2a}, depend on the pion field parametrization.  This
dependence must cancel out after summing the individual contributions associated with
these diagrams, as indeed it does~\cite{Baroni16} (this and the requirement,
remarked on below, that the axial current be conserved in the chiral limit provide
useful checks of the calculation).

The loop integrals in the diagrams of Figs.~\ref{fig:f2}--\ref{fig:f5a}
are ultraviolet divergent and are regularized in dimensional regularization~\cite{Pastore09,Pastore11,Koelling09,Koelling11,Baroni16}.
In the electromagnetic current the divergent parts of these loop integrals are reabsorbed by the LECs
$C_i^\prime$~\cite{Pastore09,Koelling11}, while those in the electromagnetic charge cancel out, in line with fact
that there are no counter-terms at N4LO~\cite{Pastore11,Koelling11}. In the case of the axial operators~\cite{ParkMinRho2,Baroni16},
there are no divergencies in the current, while those in the charge lead to renormalization
of the LECs multiplying contact-type contributions. In particular, the
infinities in loop corrections to the OPE axial charge (not shown in Fig.~\ref{fig:f5a})
are re-absorbed by renormalization of the LECs $d_i$ in the ${\cal L}^{(3)}_{\pi N}$ Lagrangian.
For a discussion of these issues we defer to Ref.~\cite{Baroni16}.

We conclude this subsection by pointing out that at the present time
two-nucleon potentials have been derived, and widely used,
up to order $(Q/\Lambda_\chi)^4$ (or $v^{(4)}$, requiring two-loop
contributions).  Very recently, a new derivation up to order
$(Q/\Lambda_\chi)^5$ has appeared~\cite{Epelbaum:2014sza}.
Some of these high-order potentials have been used, in conjunction
with the one-loop operators presented here, in calculations of
electroweak observables of light nuclei, as reported below. 
Conservation of the electromagnetic current
${\bf q}\cdot{\bf j}_\gamma=\left[\, H\, ,\, \rho_\gamma\,\right]$
with the two-nucleon Hamiltonian given by
$H=T^{(-1)}+v^{(0)}+v^{(2)}+\dots\,\,$ and
where the (two-nucleon) kinetic energy $T^{(-1)}$
is counted as $Q^{-1}$, implies~\cite{Pastore09}, order
by order in the power counting, a set of non-trivial relations
between the ${\bf j}_\gamma^{(n)}$ and the $T^{(-1)}$, $v^{(n)}$, and
$\rho_\gamma^{(n)}$ (note that commutators implicitly bring in factors
of $Q^{3}$)---incidentally, similar considerations also apply
to the conservation of the axial current in the chiral limit~\cite{Baroni16}.
These relations couple different orders in the power counting
of the operators, making it impossible to carry out a calculation,
which at a given $n$ for ${\bf j}_\gamma^{(n)}$, $v^{(n)}$, and $\rho_\gamma^{(n)}$
(and hence ``consistent'' from a power-counting perspective)
also leads to a conserved current. 

\section{Results}
\label{sec:res}

In this section we provide a sample of results obtained with $\chi$EFT
electroweak currents for light systems, including $A=2$--4 nuclei
and s- and p-shell nuclei in the mass range $A=6$--10, in the last
five years or so.  The few-nucleon calculations are based on the
chiral two-nucleon potentials developed by Entem and
Machleidt~\cite{Ent03,Machleidt11} at order $Q^4$ in the power
counting including up to two-loop corrections, and chiral three-nucleon
potentials at leading order~\cite{Epelbaum02}.  (Below, this combination
of two and three-nucleon potentials will be referred to as N3LO/N2LO,
as is customarily done in the literature, even though such a classification
does not conform to the power-counting notation adopted in
the present chapter.)  As noted earlier, the two LECs $c_D$---related
to the LEC $z_0$ in the contact axial current of Eq.~(\ref{eq:jctct})---and
$c_E$ entering the three-nucleon potential have been constrained
by fitting the Gamow-Teller matrix element in tritium $\beta$-decay
and the binding energies of the trinucleons~\cite{Gazit09,Marcucci12}.

\begin{figure}[bth]
\begin{center}
\includegraphics[height=2cm]{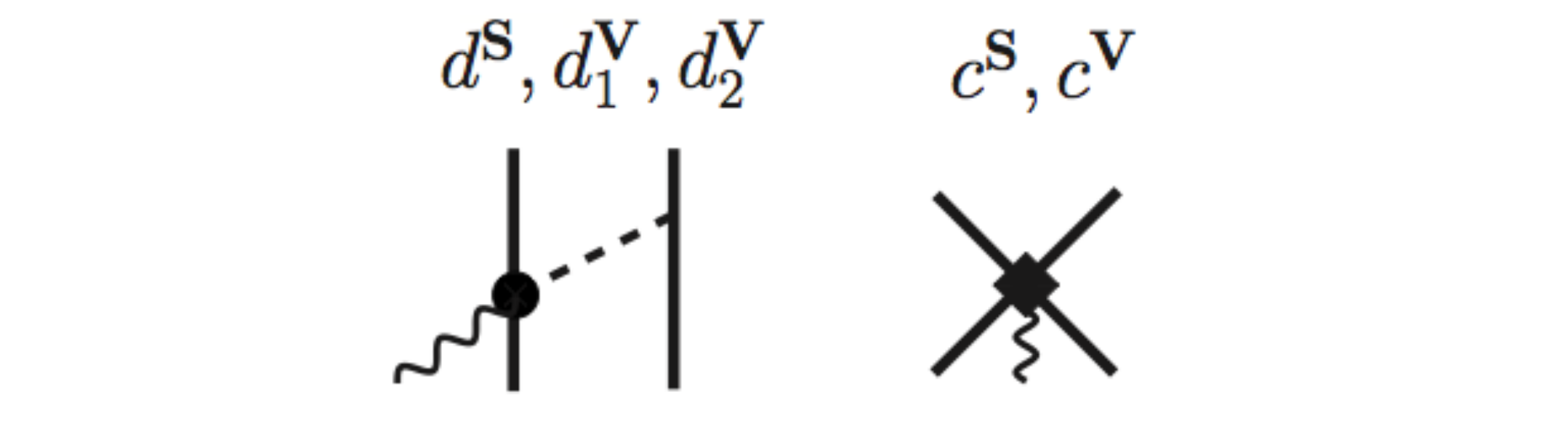}
\end{center}
\caption{The isoscalar $d^S$ and $c^S$, and isovector $d_1^V$, $d_2^V$,
and $c^V$ LECs characterizing the electromagnetic current at N3LO.}
\label{fig:fct}
\end{figure}
The electroweak operators in Figs.~\ref{fig:f2}--\ref{fig:f5a}
have power law behavior for large momenta, and need to be
regularized before they can be sandwiched between nuclear
wave functions.  The regulator is taken of the form
$C_\Lambda (k)={\rm exp}[-(k/ \Lambda)^n]$ with $n=4$
and $\Lambda$ in the range (500--600) MeV.  For processes
involving low momentum and energy transfers one would 
expect predictions to be fairly insensitive to variations of
$\Lambda$.  As shown below, this expectation is
borne out in actual calculations, at least in the case
of processes which are not inhibited at leading order,
such as the $n\, ^3{\rm He}$ radiative capture or
$p\, ^3{\rm He}$ weak fusion.

There are 5 unknown LECs in ${\bf j}$---see Fig.~\ref{fig:fct} or panels (j) and (k) of
Fig.~\ref{fig:f2}---and none in $\rho$~\cite{Pastore09,Piarulli13,Koelling09,Koelling11}.
Two (three) of these LECs multiply isoscalar (isovector) operators.
For each $\Lambda$ the two isoscalar LECs are fixed by reproducing
the deuteron and isoscalar trinucleon magnetic moments.  Two of the isovector LECs
are then constrained by assuming $\Delta$-resonance saturation~\cite{Piarulli13},
while the remaining LEC is determined by reproducing  (again for each $\Lambda$)
either the $np$ radiative capture cross section $\sigma_{np}$ at thermal neutron energies
or the isovector trinucleon magnetic moment $\mu^V$~\cite{Piarulli13}.  There are no
three-body currents entering at the order of interest~\cite{Gir10a}, and so it is
possible to use three-nucleon observables to fix some of these LECs.
Their values are listed in Table~\ref{tb:t1}.  They are generally rather large,
particularly when $c^V$ is determined by the $np$ radiative capture cross section.
The exception is the isoscalar LEC $d^S$ multiplying the one-pion exchange
current involving a sub-subleading $\gamma \pi N$ vertex from the chiral
Lagrangian ${\cal L}^{(3)}_{\pi N}$, which in a resonance-saturation
picture reduces to the $\rho\pi\gamma$ transition current.
\begin{table}
\tbl{Values for the LECs in units $1/\Lambda^2$
for $d^S$ and $1/\Lambda^4$ for $c^S$ and $c^V$; see text for further explanations.}
{\begin{tabular}{ccccc} \toprule
$\Lambda$  & $c^S$  & $d^S\times 10$ & $c^V(\sigma_{np})$ &$c^V(\mu^V)$ \\
MeV          &            &                           &                                  & \\
\hline
500 &  4.1 & 2.2 &--13  &  --8.0\\
600  & 11 & 3.2 & --22&--12 \\
\botrule
\end{tabular}}
\label{tb:t1}
\end{table}

The calculations of $A=6$--10 nuclei are carried out in the hybrid
approach, in which $\chi$EFT electroweak currents are used in
combination with he Argonne $v_{18}$ two-nucleon (AV18) and Illinois-7
three-nucleon (IL7) potentials.   The AV18 consists of a long-range component
induced by OPE and intermediate-to-short range components modeled
phenomenologically and constrained to fit the $N\!N$ database beyond
the pion-production threshold ($E_{\rm lab}=350$ MeV).  The IL7
includes a central (albeit isospin dependent) short-range repulsive term and two- and
three-pion-exchange mechanisms involving excitation of intermediate $\Delta$
resonances.  Its strength is determined by four parameters which are fixed by 
a best fit to the energies of 17 low-lying states of nuclei in the mass range $A \leq 10$,
obtained in combination with the AV18 potential.  The AV18/IL7 Hamiltonian then leads
to predictions of $\sim100$ ground- and excited-state energies up to $A=12$, including
the $^{12}$C ground- and Hoyle-state energies, in good agreement with
the corresponding empirical values (for a recent review of these as well
as results obtained for nuclear and neutron matter, see Ref.~\cite{Car15}).
 
\subsection{Electromagnetic observables of $A=2$--4 nuclei}
\begin{figure}[bth]
\includegraphics[width=11.5cm]{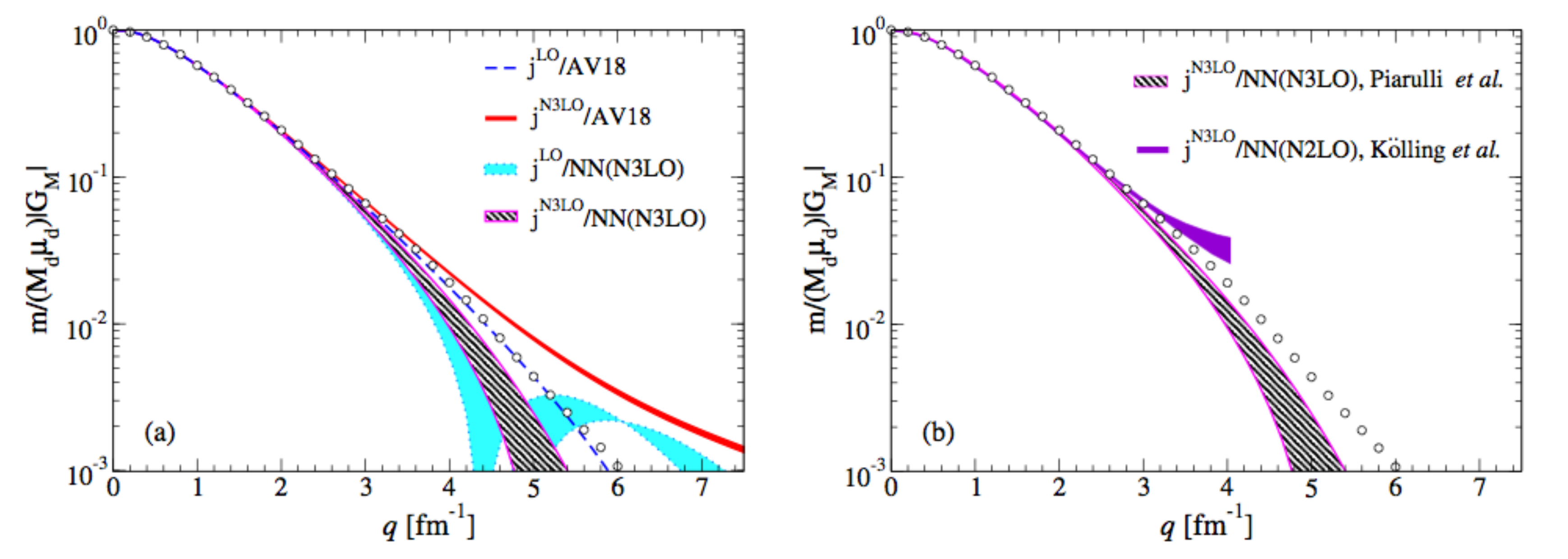}
\caption{Magnetic form factor of the deuteron: the left
panel shows results obtained with LO and N3LO currents and
either the chiral N3LO or conventional AV18 potential; the right panel shows
results obtained with N3LO currents and either the chiral N3LO
(same as in left panel) or a chiral N2LO potential by K\"olling {\it et al.}.
The bands reflect cutoff variation.  Experimental data are the empty
circles.}
\label{fig:fgm2}
\end{figure}

The deuteron magnetic form factor, calculated in
Ref.~\cite{Piarulli13}, is shown in Fig.~\ref{fig:fgm2}.
The bands reflect the sensitivity to cutoff variations in the
range $\Lambda=(500$--600) MeV.  The black bands include
all corrections up to N3LO in the (isoscalar) electromagnetic (EM) current.  The
NLO OPE and N3LO TPE currents are isovector and therefore
give no contributions to this observable.  The right panel of Fig.~\ref{fig:fgm2} contains a
comparison of the results of Ref.~\cite{Piarulli13} with those of a calculation
based on a lower order potential and in which a different strategy
was adopted for constraining the LEC's $d^S$ and $c^S$
in the N3LO EM current~\cite{Koelling12}.
This figure and the following Fig.~\ref{fig:f5} are from the
recent review paper by S.\ Bacca and S.\ Pastore~\cite{Bacca14}.

\begin{figure}[bth]
\includegraphics[width=11.5cm]{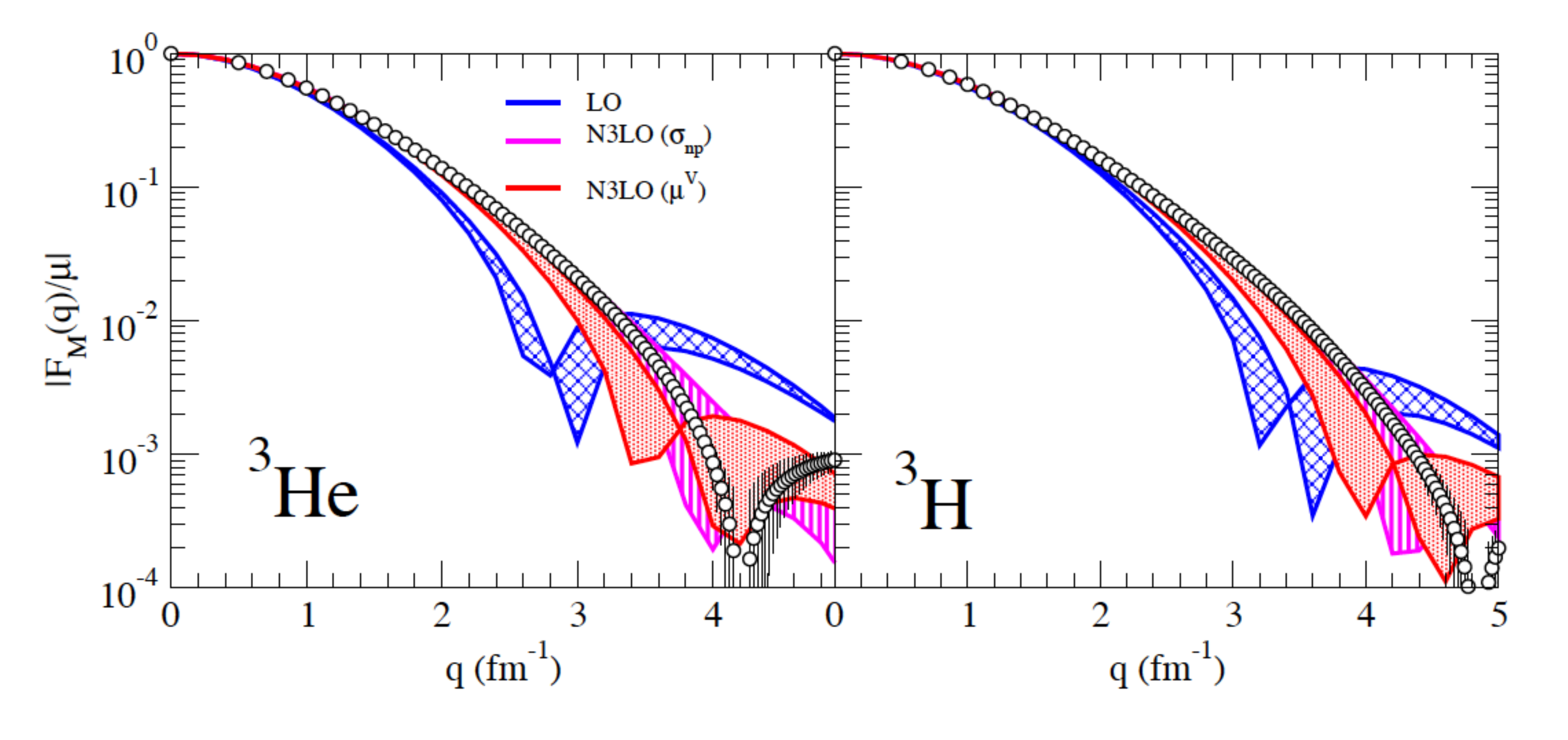}
\caption{Magnetic form factors of $^3$He (left panel) and $^3$H (right
panel); see text for further explanations.}
\label{fig:ffm3}
\end{figure}

The predicted magnetic form factors of the $^3$He and $^3$H ground states
are compared to experimental data in Fig.~\ref{fig:ffm3}~\cite{Piarulli13}.  Isovector 
OPE and TPE two-body terms in the EM current play an important role in these observables,
confirming previous results obtained in the conventional meson-exchange framework.
We show the N3LO results corresponding to the two different ways used to
constrain the LEC $c^V$ in the isovector contact current
(recall the the LECs $d_1^V$ and $d_2^V$ are assumed to be saturated
by the $\Delta$ resonance), namely by reproducing (i) the empirical value
for the $np$ cross section---curve labeled N3LO($\sigma_{np}$)---or
(ii) the isovector magnetic moment of $^3$He/$^3$H---curve labeled
N3LO($\mu^V$).  The bands display the cutoff sensitivity, which
becomes rather large for momentum transfers $q \gtrsim 3$ fm$^{-1}$.
The N3LO($\sigma_{np}$) results are in better agreement with
the data at higher momentum  transfers; however, they overestimate
$\mu^V$ by $\sim 2$\%.  On the other hand,
the N3LO($\mu^V$) results, while reproducing $\mu^V$ by construction,
under-predict $\sigma_{np}$ by $\sim 1$\% . 

\begin{figure}[bth]
\begin{center}
\includegraphics[width=5.25cm]{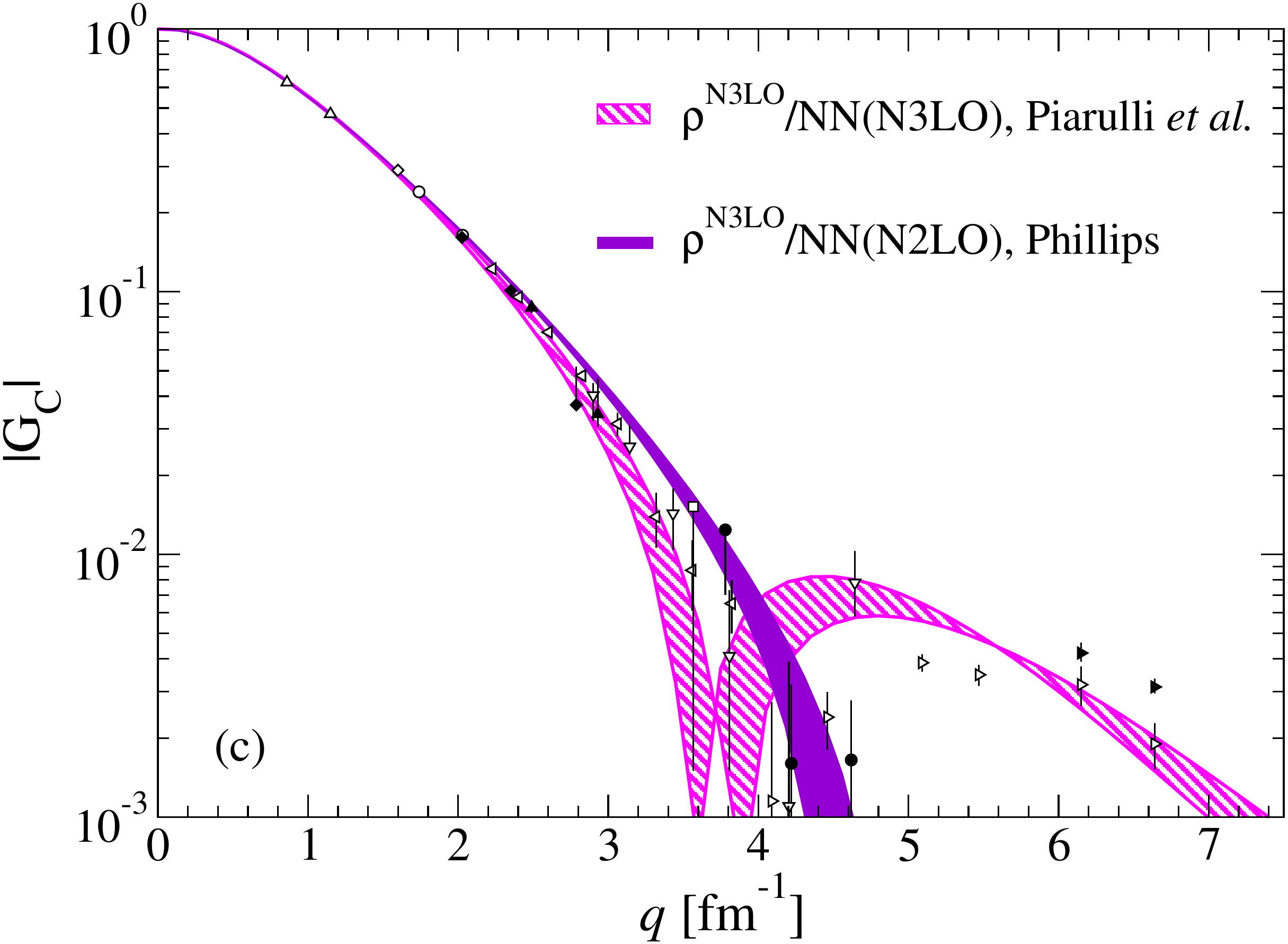}
\hspace{0.5cm}
\includegraphics[width=5.25cm]{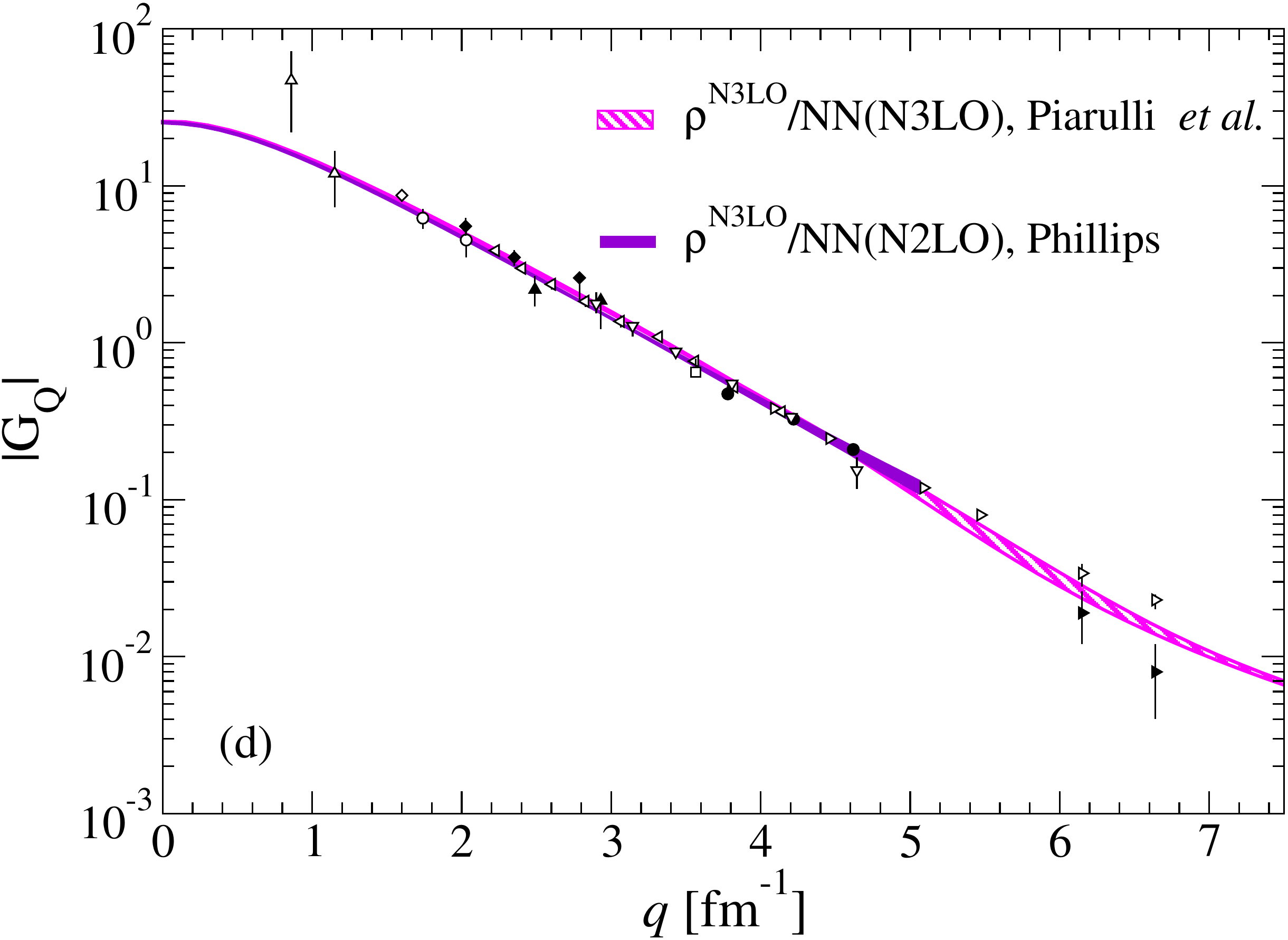}
\caption{The deuteron monopole and quadrupole form factors
obtained from measurements of the $A$ structure function and tensor polarization
are compared to predictions based on N2LO and N3LO chiral potentials.}
\label{fig:fgc2}
\end{center}  
\end{figure}

\begin{figure}[bth]
\begin{center}
\includegraphics[width=6cm]{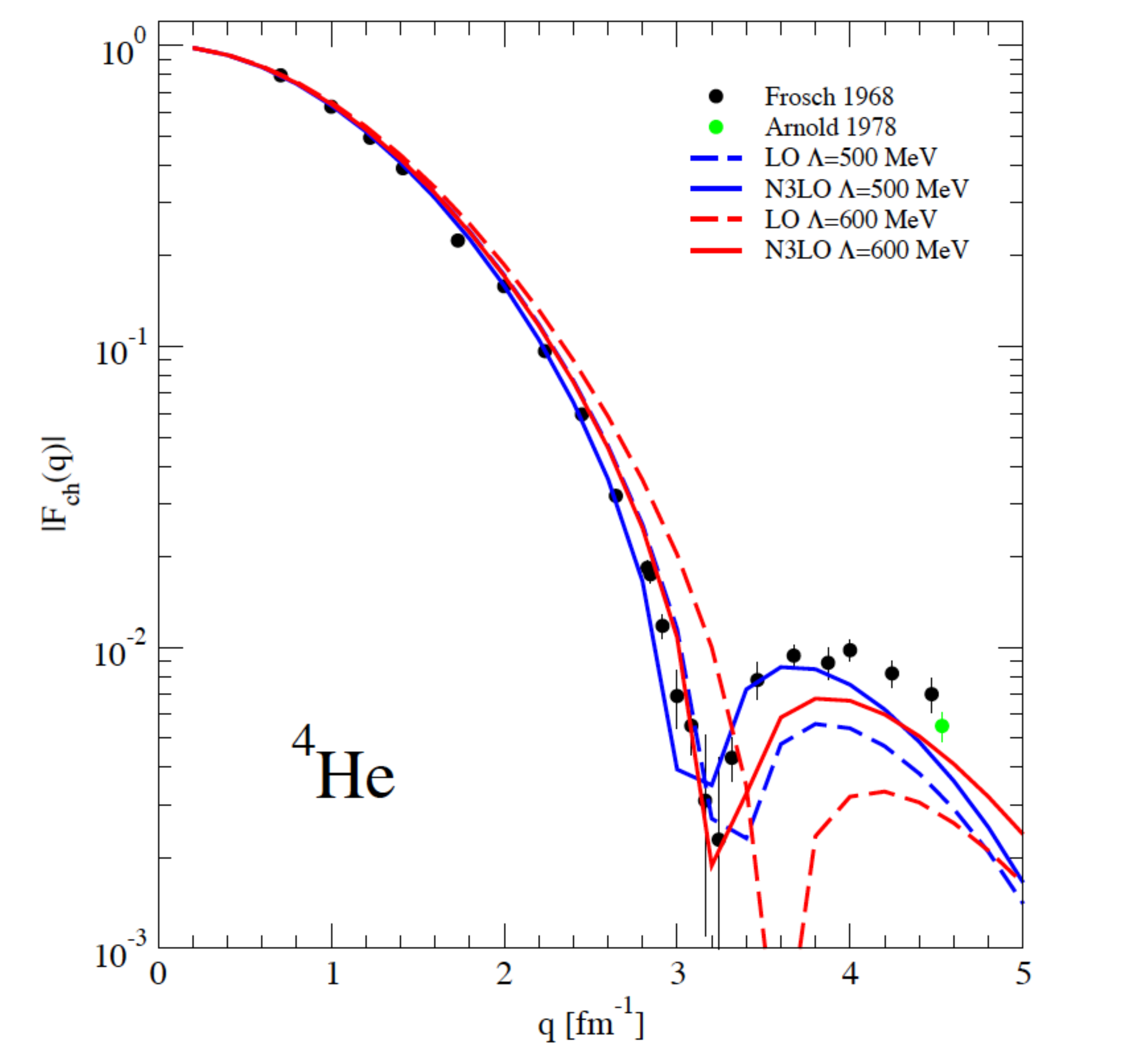}
\caption{The $^4$He charge form factor obtained from elastic
electron scattering data is compared to results obtained with the
LO and N3LO charge operator.}
\label{fig:ffc4}
\end{center}
\end{figure}

Moving on to the EM charge operator, we show in Figs.~\ref{fig:fgc2} and~\ref{fig:ffc4}
very recent calculations of the deuteron monopole and quadrupole form factors~\cite{Piarulli13} and
$^4$He (charge) form factor~\cite{Marcucci15}.  There are no unknown LECs beyond $g_A$, $f_\pi$
and the nucleon magnetic moments---the latter enter a relativistic correction,
suppressed by $Q^2$ relative to the LO charge operator, {\it i.e.}, the well-known
spin-orbit term.  The loop contributions (at N4LO) from two-pion exchange are
isovector and hence vanish for these observables.

The deuteron monopole and quadrupole form factor data are obtained
from measurements of the $A$ structure function and tensor polarization
observable in electron-deuteron scattering.  In Fig.~\ref{fig:fgc2} the
two bands correspond to two different calculations, one of which,
labeled as NN(N2LO), is based on a lower order chiral potential~\cite{Phillips03,Phillips07}.
There is good agreement between theory and experiment.
Differences between the two sets of theory predictions merely
reflect differences in the deuteron wave functions obtained with the
N3LO and N2LO potentials. These differences are amplified in the diffraction region
of the monopole form factor.

The $^4$He charge form factor is obtained from elastic electron scattering
cross section data.  These data now extend up to momentum transfers
$q \lesssim 10$ fm$^{-1}$~\cite{JLABdata}, well beyond the range of applicability of
$\chi$EFT.  In Fig.~\ref{fig:ffc4} only data up to $q\lesssim 5$ fm$^{-1}$
are shown.  They are in excellent agreement with theory.

\begin{table}[bth]
\tbl{The charge radii of the $^2$H, $^3$He, and $^4$He nuclei,
and $^2$H quadrupole moment.  The numbers in parentheses at
the side of the $\chi$EFT predictions give the cutoff dependence of
the results.}
{\begin{tabular}{ccccc} \toprule
      &  $r_c$($^2$H)  &  $Q_d$    & $r_c$($^3$He) & $r_c$($^4$He)  \\
      & (fm)                 &     (fm$^2$) & (fm)                &   (fm)    \\
\hline
$\chi$EFT           &  2.126(4)        & 0.2836(16)    &  1.962(4)  &  1.663(11)      \\
EXP & 2.130(10)  & 0.2859(6) & 1.973(14)  & 1.681(4) \\
\botrule
\end{tabular}}
\label{tb:t2}
\end{table}
Predictions for the charge radii of the deuteron and helium isotopes and for the
deuteron quadrupole moment ($Q_d$) are listed in Table~\ref{tb:t2}~\cite{Piarulli13}.  They are within
1\% of experimental values.  It is worth noting that until recently calculations
based on the conventional meson-exchange framework used to consistently
underestimate $Q_d$. However, this situation has now changed, and a
relativistic calculation in the covariant spectator theory based on a one-boson
exchange model of the $NN$  interaction has led to a value for
the quadrupole moment~\cite{Gross15} which is in agreement with experiment.

\begin{figure}[bth]
\begin{center}
\includegraphics[width=10cm]{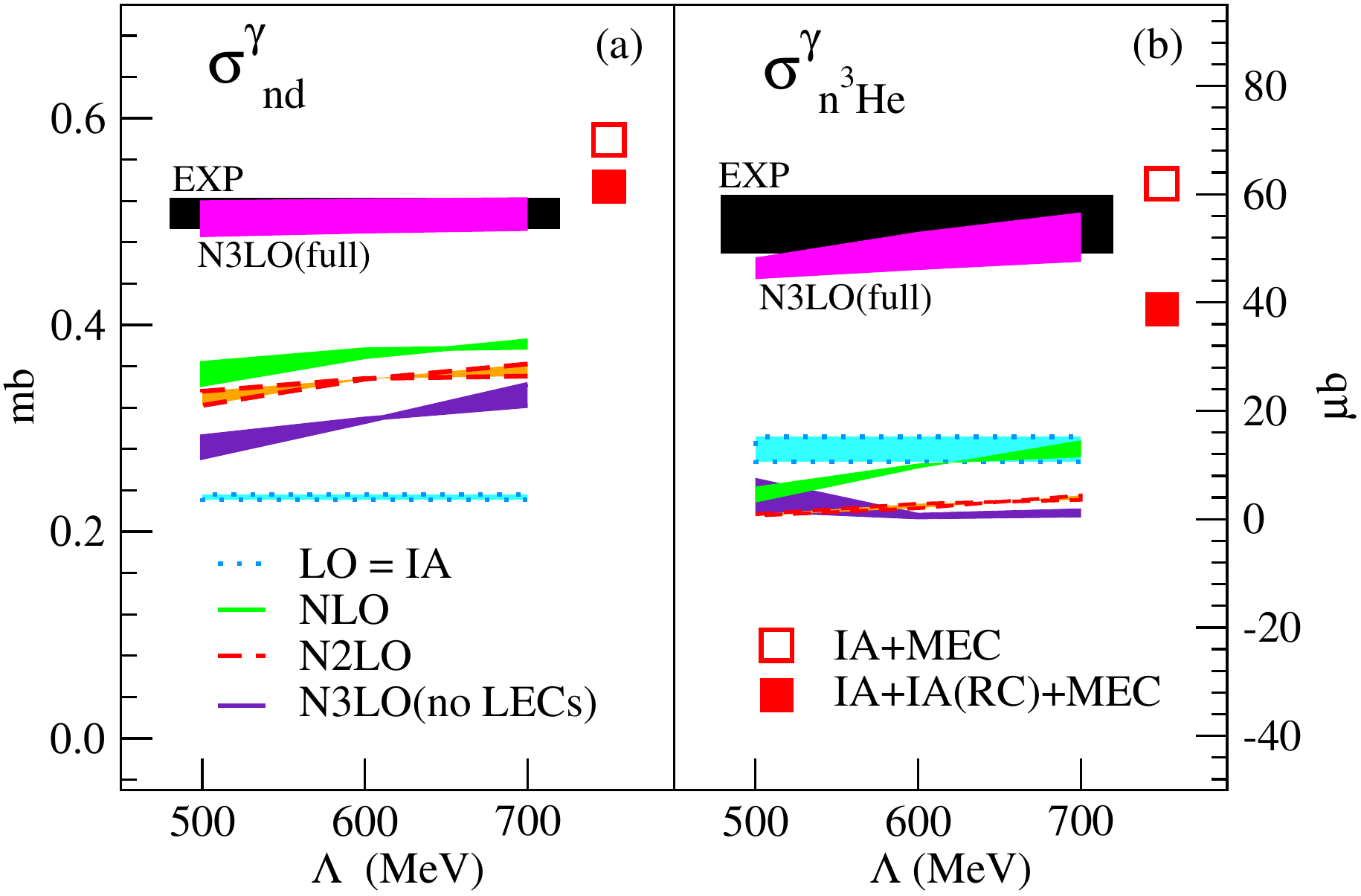}
\caption{Results for the $nd$ and $n\, ^3{\rm He}$ radiative capture
cross sections, obtained by including cumulatively the
LO, NLO, N2LO, N3LO(no LECs), and N3LO(full) contributions
from the $\chi$EFT electromagnetic current.  Also shown are
predictions obtained in the conventional approach based
on the AV18/UIX Hamiltonian and accompanying effective-meson
exchange currents, the square labeled IA+MEC and IA+IA(RC)+MEC,
the latter including relativistic correction to the IA operator.  The black
band represents the experimental data, see text for further explanations.}
\label{fig:fcap}
\end{center}
\end{figure}
As a last example we show in Fig.~\ref{fig:fcap} predictions for
the $nd$ and $n\,^3{\rm He}$ radiative capture cross sections
at thermal neutron energies~\cite{Gir10a}.   It is well known
that these $M1$ transitions are suppressed when the magnetic
dipole operator is taken to consist only of proton and
neutron contributions, i.e., in the impulse approximation (IA).
The results shown in Fig.~\ref{fig:fcap} have been obtained
from highly accurate (essentially exact) solutions of the bound and continuum
states of the $A=3$ and 4 systems with the hyperspherical-harmonics
technique~\cite{Kie08}, based on both chiral and conventional two-
and three-nucleon potentials, the N3LO/N2LO and AV18/UIX,
and chiral electromagnetic currents up to N3LO.  However,
in this earlier study, the procedure adopted to fix the three isovector
LECs is different from that utilized in the calculations discussed
so far.  Here $\Delta$-resonance saturation is exploited only to fix
the ratio of the two LECs $d_1^V$ and $d_2^V$ in the tree-level
contribution of Fig.~\ref{fig:fct}, and the remaining two (isovector)
LECs are then determined by a simultaneous fit to $\sigma_{np}$ and
the isovector combination of the trinucleon magnetic moment.
Furthermore, the LECs in the contact current originating from
minimal couplings have been taken from a lower order (NLO)
chiral potential~\cite{Pastore09} rather than from the
N3LO potential of Ref.~\cite{Ent03,Machleidt11}.

In Fig.~\ref{fig:fcap} the experimental data
are from Ref.~\cite{Jurney82} for $nd$ and
Refs.~\cite{Wolfs89,Werve91} for $n\, ^3$He, the band thickness
denoting the error.  Results obtained with the
complete N3LO $\chi$EFT operator are shown
by the orange band labeled N3LO(full):  those corresponding
to the N3LO/N2LO (AV18/UIX) model delimit the lower
(upper) end of the band in the case of $nd$, and its
upper (lower) end in the case of $n\, ^3{\rm He}$.
There is considerable cutoff dependence, particularly
for the four-body capture.  In this connection, it is
interesting to note the crucial role played by the
N3LO currents in Fig.~\ref{fig:fct}: indeed, retaining
only the minimal contact currents and the
currents from TPE loop
corrections---bands labeled
N3LO(no LECs)---would severely under-estimate the
measured cross sections.  It is clear that the convergence
of the chiral expansion for these processes 
is problematic.  The LO (or IA) is unnaturally
small, since the associated operaotr cannot
connect the dominant S-states in the
hydrogen and helium bound states.
 This leads to an enhancement of the NLO
 contribution, which, however, in the case of
 $n\, ^3{\rm He}$, is offset by the destructive interference between
 it and the LO contribution.  Thus a satisfactory
 description of these processes remains particularly
 challenging for nuclear theory and nuclear $\chi$EFT
 in particular.

\subsection{Electromagnetic transitions in $A=6$--10 nuclei}

Heavier systems offer new challenges and opportunities for
applications of nuclear $\chi$EFT, and {\it ab initio} studies
of electroweak processes based on this approach in systems
with $A > 4$ have only just begun.  In the mass range $A=6$--12
Variational Monte Carlo (VMC) and Green's function Monte
Carlo (GFMC) methods allow us to carry out accurate, in fact exact
in the case of GFMC, first-principles calculations of many nuclear
properties (see Ref~\cite{Car15} for a recent review).  However,
since these methods are formulated in configuration space, it has
not been possible to use them in conjunction with the chiral
two-and three-nucleon potentials above, which are given in
momentum-space (and are strongly non-local in configuration space).
A first step in this direction is the very recent development of a
class of configuration-space, minimally non-local two-nucleon
chiral potentials that fit the $np$ and $pp$ database up to the
pion-production threshold with a $\chi^2$ per datum close to
1.3~\cite{Piarulli15}, i.e., of the same quality as the well established
N3LO models of Refs.~\cite{Ent03,Machleidt11}.  Use of these
potentials in VMC and, especially, GFMC calculations of light s- and p-shell
nuclei will expand the scope of the nuclear $\chi$EFT approach,
and in particular test its validity beyond the realm of few-nucleon
systems, to which it has primarily been limited so far. 

\begin{figure}[bth]
\begin{center}
\includegraphics[width=10cm]{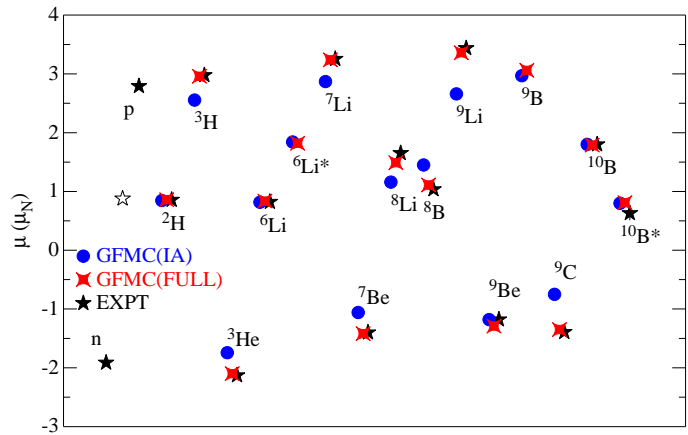}
\caption{Magnetic moments in nuclear magnetons for $A \leq 10$ nuclei
obtained in GFMC calculations based on the conventional AV18/IL7 Hamiltonian
and chiral electromagnetic currents.  Black stars indicate the experimental
values, while the blue (red) dots represent the results of calculations
including the LO (N3LO) chiral electromagnetic current. }
\label{fig:fmu}
\end{center}
\end{figure}
\begin{figure}[bth]
\begin{center}
\includegraphics[width=8cm]{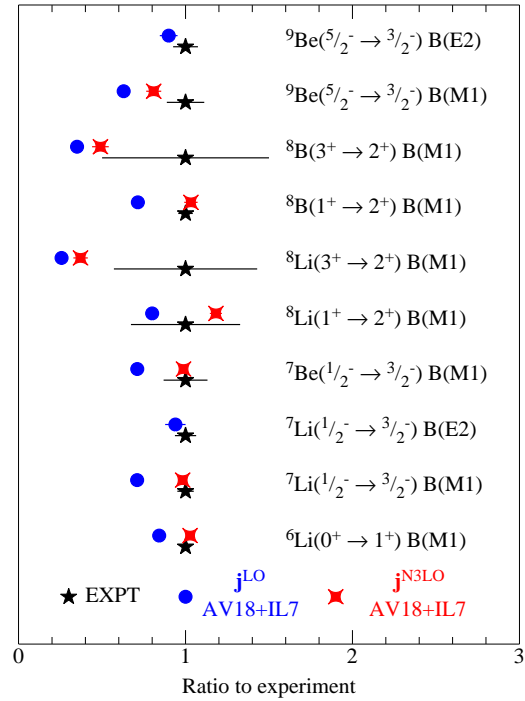}
\caption{Ratio to the experimental $M1$ and $E2$ transition widths
in $A \leq 9$.  Black stars with error bars indicate the experimental
values, while the blue dots (red diamonds) represent GFMC AV18/IL7
calculations including chiral electromagnetic currents at LO (up to N3LO). }
\label{fig:fb1}
\end{center} 
\end{figure}

As mentioned earlier, the results presented in this subsection
for the magnetic moments and transition
widths of $A=6$--10 nuclei~\cite{Pastore13} have been obtained
in the hybrid approach, which combines
conventional potentials (AV18 and IL7) with chiral
electromagnetic currents, see Figs.~\ref{fig:fmu} and~\ref{fig:fb1}.
Figure~\ref{fig:fmu} makes it plain
that the inclusion of corrections beyond LO is necessary in order
to have a satisfactory description of the experimental data:
their effect is particularly pronounced in the $A=9$ and isospin
$T=3/2$ systems , in which they provide up to $\sim 20$\%
(40\%) of the total predicted value for the $^9{\rm Li}$ ($^9{\rm C}$)
magnetic moments.

In Fig.~\ref{fig:fb1} the calculated $M1$ and $E2$ transition widths for
$A=6$--9 nuclei are compared to experimental data.  Overall,
there is good agreement between theory and experiment, particularly
when one considers the fact that for systems like $^8$Li and
$^8$B the errors bars are so large to prevent any robust conclusions
to be drawn from the apparent under-prediction
by theory of the associated widths.   It should be noted
that the $E2$ widths have been obtained with the LO EM
charge operator.  Higher order corrections
are expected to be very small, in particular 
the OPE charge operator in Eq.~(\ref{eq:pich}) vanishes in the static limit.
\subsection{Weak transitions in few-nucleon systems}
Most calculations of nuclear axial current matrix elements, such as those discussed
below for the $pp$ weak fusion and for muon capture on $^2$H and
$^3$He, have used axial current operators up to N3LO or $Q^0$ (one exception is
Ref.~\cite{Klos13}, which included effective one-body reductions, for use in a
shell-model study, of some of the TPE corrections to the axial current
derived in Ref.~\cite{Parketal}).  A recent application of these N3LO transition operators
is the calculation of the rates for $\mu^{-}$ capture on deuteron and $^3$He~\cite{Marcucci12}.
These rates have been predicted with $\sim 1$\% accuracy,
\[
\Gamma(^2{\rm H})=(399 \pm 3)\, {\rm sec}^{-1} \ , \qquad \Gamma(^3{\rm He})=(1494 \pm 21)\, {\rm sec}^{-1} \ .
\]
At this level of precision, it is necessary to also account for electroweak radiative corrections,
which have been evaluated for these processes in Ref.~\cite{Cza07}.  The error quoted on
the predictions above results from a combination of (i) the experimental error
on the $^3$H GT matrix element used to fix the LEC in the contact axial current,
 (ii) uncertainties in the electroweak radiative corrections---overall, these corrections
 increase the rates by 3\%---and (iii) the cutoff dependence.
 
There is a very accurate and precise measurement of the rate
on $^3$He: $\Gamma^{\rm EXP}(^3{\rm He})=(1496\pm 4)$ sec$^{-1}$~\cite{Ack98}.
It can be used to constrain the induced pseudo-scalar
form factor of the nucleon. It gives $G_{PS}(q_0^2=-0.95\, m_\mu^2)=8.2 \pm 0.7$,
which should be compared to a direct measurement on hydrogen at PSI,
$G^{\rm EXP}_{PS}(q_0^2=-0.88\, m_\mu^2)=8.06 \pm 0.55$~\cite{Mucap},
and a chiral perturbation theory prediction of $7.99 \pm 0.20$~\cite{Bern94,Kai03}.

The situation for $\mu^-$ capture on $^2$H remains, to this day, somewhat confused:
there is a number of measurements that have been carried out, but
they all have rather large error bars.  However, this unsatisfactory state of affairs
should be cleared by an upcoming measurement of this rate by the MuSun collaboration at PSI
with a projected 1\% error.

Another recent example is the proton weak capture on protons~\cite{Parketal,Marcucci13}.
This process is important in solar physics: it is the largest source of
energy and neutrinos in the Sun.  The astrophysical $S$-factor
for this weak fusion reaction is one of the inputs in the standard
model of solar (and stellar) evolution~\cite{Bahcall04}.  A recent calculation based
on N3LO chiral potentials including a full treatment of
EM interactions up to order $\alpha^2$ ($\alpha$ is
the fine structure constant), shows that
it is now predicted with an accuracy of much less than 1\%:
$S(0)=(4.030 \pm 0.006) \times 10^{-23}$ MeV-fm$^2$.
\begin{figure}[bth]
\begin{center}
\includegraphics[height=6cm,width=9cm]{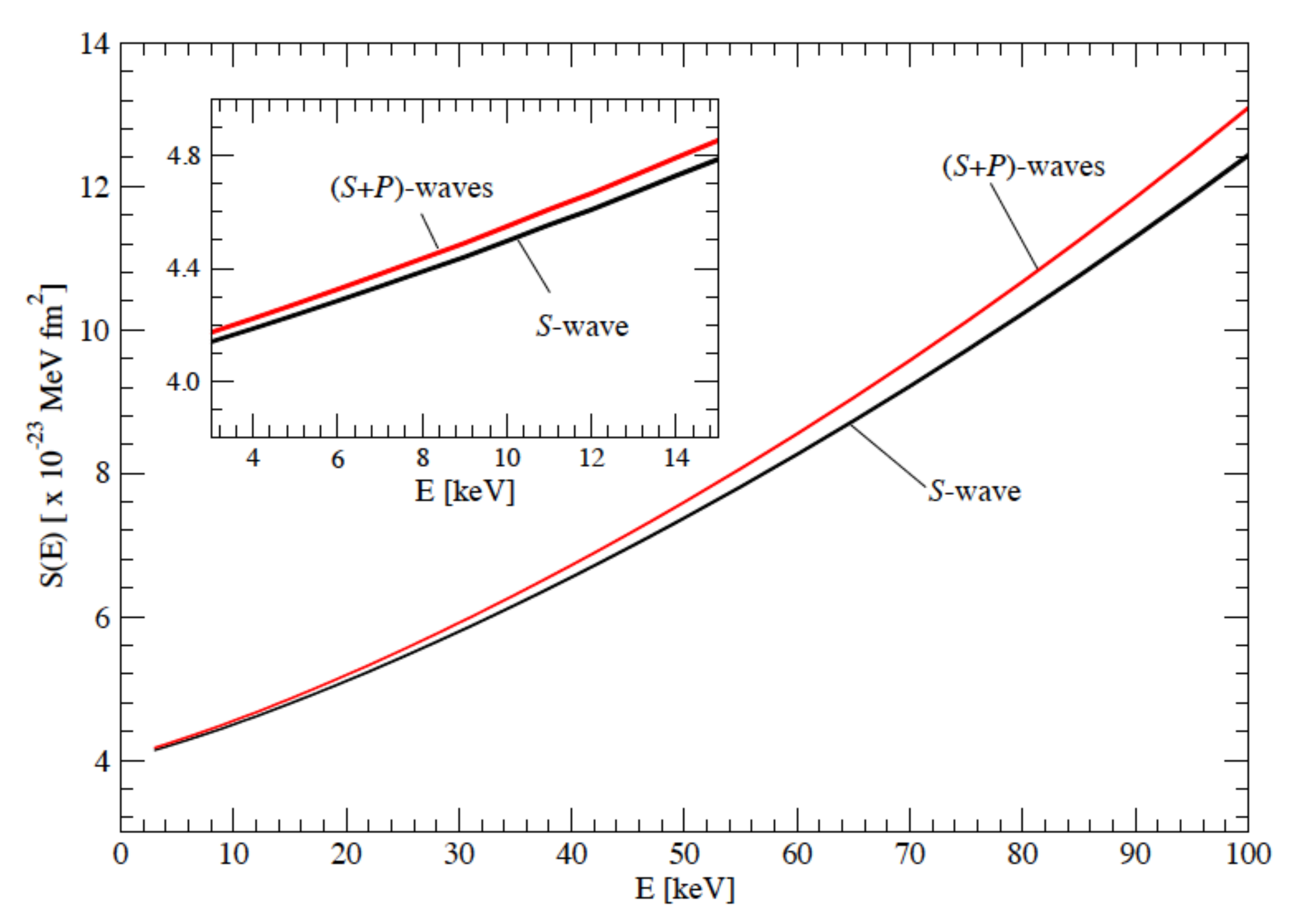}
\caption{The $S$-factor for $pp$ weak fusion due to S- and (S+P)-wave
capture as function of energy.}
\label{fig:f7}
\end{center}
\end{figure}
This calculation also included the (small) effects from
capture of the two protons in relative P-wave, see Fig.~\ref{fig:f7}~\cite{Marcucci13}.
The increase due to P-wave capture offsets the decrease
from higher order EM effects, in particular vacuum polarization.
\section{Conclusion}
The presentation above illustrates the remarkable progress of the development and
application of the chiral Lagrangian based description of nuclear electroweak current
operators beginning with the first steps taken by Gerry Brown, Mannque Rho and
their colleagues and students around 1970. The early work in the 1970s was, however,
based on the lowest order terms in the chiral Lagrangians and, to a large extent,
phenomenological wave functions. The advent of systematic chiral effective field
theory has brought the theoretical work to a quantitative level and has
provided a basis for it in the fundamental theory of the strong interactions.

\vspace{0.5cm}
\noindent We wish to thank our collaborators A.\ Baroni, J.\ Carlson, L.\ Girlanda, A.\ Kievsky,
L.E.\ Marcucci, S.\ Pastore, M.\ Piarulli, S.C.\ Pieper, M.\ Viviani, and R.B.\ Wiringa
for their many contributions to the work presented here.  The support of the U.S.~Department
of Energy under contract DE-AC05-06OR23177 is also gratefully acknowledged.

\bibliographystyle{ws-rv-van}
\bibliography{mec_riska.bib,FF-review_pma.bib,FF-review_chieft.bib,FF-review_intro.bib,FF-review_res.bib,bib-CST-AT.bib}
\end{document}